\documentclass{article}
\usepackage[preprint, nonatbib]{nips}

\usepackage{amsmath,amssymb,amsfonts}
\usepackage{algorithmic}

\usepackage{graphicx}
\usepackage{epstopdf}

\usepackage{textcomp}
\usepackage{xcolor}
\usepackage{mathtools}
\usepackage{comment}
\usepackage{caption}
\usepackage{subcaption}
\usepackage{tabularx,booktabs}
\usepackage{slashbox}
\usepackage{booktabs,caption}
\usepackage[flushleft]{threeparttable}
\usepackage{multirow}

\usepackage{amsmath}
\usepackage{color}
\usepackage{comment}

\usepackage{mathtools}
\usepackage{amsthm}

\usepackage{overpic}

\usepackage{tabularx} 
\usepackage{wrapfig}
\usepackage{wrapfig, blindtext}
\usepackage{cases}
\usepackage{empheq}
\usepackage{url}
\usepackage{float}

\newcommand{\defeq}{\vcentcolon=}
\DeclareMathOperator{\kl}{KL}

\newtheorem{theorem}{Theorem}

\newtheorem{proposition}[theorem]{Proposition}

\newcommand\norm[1]{\lVert#1\rVert}
\newcommand\bignorm[1]{\big\lVert#1\big\rVert}

\usepackage{tablefootnote}
\usepackage{footnote}

\newcolumntype{C}[1]{>{\centering\arraybackslash}p{#1}}

\usepackage{minitoc}

\usepackage{wrapfig}
\usepackage{wrapfig, blindtext}
\usepackage{csquotes}

\usepackage{soul}
\usepackage{xcolor}
\newcommand{\rankone}[1]{\sethlcolor{black!50}\hl{#1}}
\newcommand{\ranktwo}[1]{\sethlcolor{black!20}\hl{#1}}

\DeclareMathOperator*{\argmin}{arg\,min}

\author{
Tian Guo \,\,\,\,\,\,\,\, Emmanuel Hauptmann
  \\
  Systematic Equities Team,
  RAM Active Investments
  \\
  Geneva, Switzerland 
  \\
  \texttt{ \{tig, eh\}@ram-ai.com }
}

\begin{document}

\title{
Exploring the Synergy of Quantitative Factors and Newsflow Representations from Large Language Models for Stock Return Prediction 
}
\maketitle

\doparttoc 
\faketableofcontents 

\begin{abstract}
In quantitative investing, return prediction supports various tasks, including stock selection, portfolio optimization, and risk management. 
Quantitative factors, such as valuation, quality, and growth, capture various characteristics of stocks.
Unstructured data, like news and transcripts, has attracted growing attention, driven by recent advances in large language models (LLMs).
This paper examines effective methods for leveraging multimodal factors and newsflow in return prediction and stock selection.
First, we introduce a fusion learning framework to learn a unified representation from factors and newsflow representations generated by an LLM.
Within this framework, we compare three methods of different architectural complexities: representation combination, representation summation, and attentive representations.
Next, building on the limitation of fusion learning observed in empirical comparison, we explore the mixture model that adaptively combines predictions made by single modalities and their fusion.
To mitigate the training instability of the mixture model, we introduce a decoupled training approach with theoretical insights.
Finally, our experiments on real investment universes reveal:
(1) Within fusion learning, the representation combination method, despite its relatively low architectural complexity, generally outperforms other fusion methods.
The relative performance of fusion methods varies across universes, suggesting that the predictive relevance of data, particularly news, may differ across markets.
(2) The mixture model appears relatively robust and balanced across universes and portfolios, delivering comparable or superior performance.
Its enhanced adaptability can be beneficial in settings where the predictive relevance of factors and news is likely more variable.
(3) Fine-tuning the LLM during the training of these multimodal models does not consistently benefit performance; its impact varies across universes, potentially reflecting differences in market efficiency and data characteristics.
%
%
%
%
%
%
\end{abstract}

\section{Introduction}

Quantitative investing involves using numerical features, also referred to as quantitative factors in finance, derived from diverse data sources (e.g., prices, economic indicators, and analyst estimates), to select stocks and construct portfolios~\cite{fama1996multifactor, ang2014asset}.
Traditional quantitative factors, e.g., value, momentum, and growth, have demonstrated predictive power for market movements in numerous studies~\cite{gu2020empirical, chauhan2020uncertainty, duan2022factorvae}.
Recently, the incorporation of textual data, such as financial news, earnings call transcripts, and annual reports, has gained significant traction, driven by advances in large language models (LLMs)~\cite{qin2019you, sawhney2020deep, lopez2023can, rahimikia2024r}.

\begin{figure}[t]
  \begin{center}
    \includegraphics[width=0.95\textwidth]{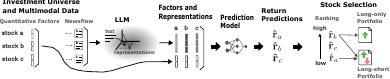}
  \end{center}
  \caption{
  Workflow using quantitative factors and newsflow for return prediction and stock selection.
  }
  \label{fig:synergy_workflow}
\end{figure}

This paper focuses on predicting stock returns using multimodal quantitative factors and financial newsflow, as illustrated in Fig.~\ref{fig:synergy_workflow}.
Accurate return forecasting is essential for subsequent tasks like stock selection and portfolio optimization~\cite{gu2020empirical, chen2024deep}.
Quantitative factors, grounded in financial theory, capture fundamental aspects of stocks, while financial news provides timely information about company events and actions.
These two modalities offer complementary perspectives, making their integration promising for prediction tasks~\cite{ramachandram2019deep, zou2022stock, yuan2025survey}.

While quantitative factors and financial newsflow have each been studied separately in existing works~\cite{chauhan2020uncertainty, duan2022factorvae, guo2024fine, rahimikia2024r}, combining them poses several challenges.
First, the two modalities differ fundamentally in structure: quantitative factors are structured and numeric, while newsflow is unstructured and textual.
Second, their predictive relevance, that is, the degree to which a modality provides predictive power for stock returns, can vary.
News, in particular, is context-dependent and may often be less relevant or provide little incremental information relative to factors that already capture (or price in) various aspects of stocks~\cite{hu2018listening, wang2024modeling, duan2022factorvae, gu2020empirical}.
Even after filtering out irrelevant news, which remains a non-trivial task in practice, the relative predictive relevance of news versus factors can still fluctuate depending on data characteristics and market conditions.

\textbf{Contributions.} 
This paper investigates effective model designs and training methods for utilizing quantitative factors and newsflow in return prediction and stock selection.
Our contributions are:

(1) We formulate the problem of return prediction and stock selection using multimodal factors and newsflow.
We introduce a multimodal fusion learning framework to learn unified representations by combining factors with newsflow representations generated by an LLM.
In this framework, we compare three methods of distinct architectural complexities: representation combination, representation summation, and attentive representation.

(2) Motivated by empirical comparison in fusion learning, we explore the mixture model that adaptively weights the predictions made by single modalities and their fusion.
This enables information integration at both the representation and prediction levels.
We observe that conventional training of such mixture models often leads to instability and performance degradation. 
To mitigate this issue, we provide theoretical insights into the cause and introduce a decoupled training method.

(3) We conduct experiments on real data across multiple investment universes.
We build long-only and long-short portfolios using return predictions and evaluate their backtest performance.
In addition, we compare results obtained without and with LLM fine-tuning during the training of prediction models.
These experiments yield several insights into effective multimodal modeling of factors and news for stock return prediction, which are summarized in the Conclusion section.
%
%


%
%


\section{Related Work}


\textbf{Multimodal Learning for Finance.}
Integrating multimodal data has gained increasing attention in finance~\cite{antulov2021temporal, bhatia2024fintral, zhang2024multimodal}.
\cite{bhatia2024fintral, huang2024open} developed multimodal financial LLMs for various analytical tasks.
For prediction tasks, social media and event data were used with numerical features to capture sentiment and event-driven price dynamics~\cite{weng2018predicting, xu2018stock, wang2019ean}.
Some works developed graph and attention-based methods to fuse multimodal data for predicting volatilities and earnings~\cite{ang2022guided, koval2024financial}.
This paper presents a comparative study of different multimodal fusion methods for return prediction and stock selection.


\textbf{LLMs in Quantitative Investment.}
Previous works used word-level embedding techniques for modeling stock and forex movements~\cite{liu2018hierarchical, hu2018listening, chen2019group}.
Recent advances in LLMs have significantly improved contextual understanding and can generate powerful numeric representations of text for prediction tasks~\cite{tang2024understanding, guo2024fine, songomnipred2024}.
\cite{araci2019finbert, liu2021finbert, iacovides2024finllama} fine-tuned pre-trained LLMs for financial sentiment extraction.
\cite{lopez2023can, wang2024llmfactor, koa2024learning, tong2024ploutos} employed prompt-based methods to harness the reasoning capabilities of LLMs over financial data.
\cite{kim2024financial} utilized chain-of-thought prompts~\cite{wei2022chain} to analyze financial statements.
\cite{rahimikia2024r} trained LLMs using temporally split datasets to mitigate look-ahead bias.
\cite{li2024alphafin} used retrieval-augmented data to improve financial analysis.
\cite{chen2022expected} developed the sentiment and return prediction models with LLM-generated text representations of news. 

In this paper, we use the newsflow representations generated by an LLM in fusion learning and mixture modeling.
Meanwhile, we compare the performance with and without fine-tuning the LLM during the training of our multimodal prediction models~\cite{hu2021lora}.


\textbf{Mixture Models for Financial Predictions.}
Mixture models enable adaptive learning to combine multiple specialized components~\cite{yuksel2012twenty, shi2019variational, antulov2021temporal}.
\cite{sun2023mastering} trained mixtures of stock return prediction components using price-based features, but did not indicate the specialization of each mixture component.
\cite{ding2024tradexpert} developed a mixture of LLMs, i.e., a separate LLM for each type of financial data and an additional LLM for aggregating the predictions, without the need for a model training process.
\cite{saqurfiltered} combined several pre-trained expert LLMs through filtering for online time-series prediction tasks.

The mixture model in this paper involves prediction components that correspond to single modalities and their fusion.
Moreover, we identify instability in the conventional training of this mixture model, analyze its causes, and introduce a specialized training method with theoretical insights.



\section{Factors and Newsflow for Return Prediction and Stock Selection}


\subsection{Problem Statement}\label{sec:problem}
We consider an investment universe consisting of a set of stocks denoted by \( \mathcal{U} = \{ s \}_{s=1}^S \), where each \( s \) represents a stock index.
The prediction target is the \( \ell \)-step forward return of stock \( s \) at time \( t \), denoted by \( r_{s,t + \ell} \in \mathbb{R} \).
For a target \( r_{s,t + \ell} \), we define the corresponding vector of quantitative factors of stock $s$ at timestamp $t$ as \( \mathbf{x}_{s,t,f}  \in \mathbb{R}^{d_f} \).

Meanwhile, we use stock-specific news, referring to news reporting events related to a company (e.g., earnings releases, management changes, product launches).
A news item published at time $i$ for stock $s$ is denoted by $\mathbf{N}_{s, i}$.
To predict \( r_{s,t + \ell} \), we collect the news in a look-back window before time $t$, forming the newsflow $\{ \mathbf{N}_{s, i} \}_{ i \in \mathcal{T}_{s, <t} }$ where $\mathcal{T}_{s, <t}$ represents the set of relevant timesteps.

Following prior works~\cite{guo2024fine, tang2024understanding, behnamghader2024llm2vec}, we adopt a simple approach without trainable parameters to obtain the newsflow representation $\mathbf{x}_{s, t, n}\in\mathbb{R}^{d_n}$:
feeding $\{ \mathbf{N}_{s, i} \}_{ i \in \mathcal{T}_{s, <t} }$ into an LLM and aggregating the resulting token representations into one vector $\mathbf{x}_{s, t, n}$.
The LLM can be fine-tuned by backpropagating through $\mathbf{x}_{s, t, n}$ when training the prediction model, as illustrated in Fig.~\ref{fig:synergy_methods}.

The training data is formed by collecting instances across stocks and timestamps, denoted as $\mathcal{D} \defeq \{ ( \mathbf{x}_{s,t,f}, \mathbf{x}_{s,t,n}, r_{s,t + \ell} ) \}_{s \in \mathcal{U}, t \in \mathcal{T}}$, where \( \mathcal{T} \) represents the timestamps in the training period.
For simplicity, we omit the indices $s$ and $t$ in the remainder of the paper and denote a generic instance sample as $\{\mathbf{x}_f, \mathbf{x}_n, r\} \sim \mathcal{D}$.

At test time, we evaluate an important application of return predictions: selecting stocks into Long-Only and Long-Short portfolios and backtesting their performance~\cite{gu2020empirical, chauhan2020uncertainty}, as illustrated in Fig.~\ref{fig:synergy_workflow}.

\underline{Long-Only Portfolios} include stocks with the expectation of the highest forward return.
In practice, it is built by ranking the stocks based on return predictions and selecting the top-K stocks.
$K$ is usually chosen according to the decile or quantile of the universe, e.g., $10\%$ of the number of stocks.

\underline{Long-Short Portfolios} include stocks with the highest and lowest return expectations.
The stocks with the lowest returns are expected to experience a price drop, and the portfolio can profit by selling them at the current price and repurchasing them at a lower price in the future.
It is built by including the top-K and bottom-K stocks based on return predictions.


\subsection{Methodologies}
\begin{wrapfigure}[14]{r}{0.59\textwidth}
  \begin{center}
    \includegraphics[width=0.59\textwidth]{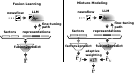}
  \end{center}
  \caption{
  Illustration of fusion learning and mixture model.
  }
  \label{fig:synergy_methods}
\end{wrapfigure}

In this section, we explore two categories of approaches as illustrated in Fig.~\ref{fig:synergy_methods}.
First, from a multimodal learning perspective, it is essential to obtain a unified representation that effectively integrates information from different modalities~\cite{zou2022stock, zhao2024deep, yuan2025survey}.
Accordingly, we present a representation-level fusion learning framework that combines factors and newsflow representations into a unified representation for return prediction.

However, while factors grounded in financial theories tend to offer relatively stable predictive power~\cite{gu2020empirical}, news data is inherently noisy, and its predictive relevance depends on its content and the information it provides beyond factors~\cite{wang2024modeling}.
Although fusion learning can leverage attention mechanisms to weight modalities, it ultimately predicts based on unified representations, thereby entangling factor and news information.
This lack of explicit separation of different predictive relevance can undermine performance, as illustrated in Fig.~\ref{fig:na_obs_perf}.

Subsequently, we explore the mixture model that adaptively combines predictions separately generated from factors and unified representations.


\subsubsection{Fusion Learning over Factors and Newsflow Representations}\label{sec:fusion_learn}
Our fusion learning framework comprises two functions: a representation fusion function and a prediction function, as formulated in Eq.~\ref{eq:fusion_framework}: 
\begin{align}
& 
\hat{r} = g_{}( \mathbf{x}_u )
\,\,\,\,\,\,
\mathbf{x}_u = z( \mathbf{x}_f, \mathbf{x}_n )
\label{eq:fusion_framework} 
\\
& 
\min_{\theta_u} \mathbb{E}_{ \{\mathbf{x}_f, \mathbf{x}_n, r\} \sim \mathcal{D} } 
\big[  ( r - \hat{r} )^2 \big] 
\label{eq:fusion_loss} 
\end{align}
, where $z(\cdot, \cdot)$ denotes the fusion function that integrates the factors and newsflow representations into a unified representation $\mathbf{x}_u$.
$g(\cdot)$ is the prediction function mapping $\mathbf{x}_u$ to the predicted return $\hat{r}$.
The trainable parameters $\theta_u$, including those in both the fusion and prediction functions, are optimized via stochastic gradient descent-based optimization to minimize the expected squared error between the predicted return $\hat{r}$ and the true value $r$ as Eq.~\ref{eq:fusion_loss}.
Within this framework, different fusion strategies can be instantiated by specifying the form of $z(\cdot, \cdot)$.

Next, we present three representative methods that span a range of architectural complexities, from simple dense layer-based to attention-based fusion (with implementation details in the Appendix).


\textbf{Representation Combination.}
By treating each dimension of the newsflow representation as a feature alongside the numerical factors, combinations of all these features form a unified representation~\cite{moon2024anymal}.
A straightforward approach is to concatenate the two and pass through a dense layer that learns arbitrary nonlinear weighted combinations of input features~\cite{chen2020uniter, zhang2025camef}, as Eq.~\ref{eq:rep_combine}:
%
\begin{align}\label{eq:rep_combine}
z( \mathbf{x}_f, \mathbf{x}_n ) = h( \mathbf{x}_f  \oplus \mathbf{x}_n ) 
\end{align}
, where $h(\cdot)$ represents a dense layer, and $\oplus$ denotes the concatenation operation.


\textbf{Representation Summation.} 
By assuming a shared representation space, the Representation Summation method projects each modality into a vector of equal dimensionality and then sums these projected representations~\cite{kiela2018efficient, wang2022ofa}, thereby encouraging modality alignment, as shown in Eq.~\ref{eq:rep_sum}:
\begin{align}\label{eq:rep_sum}
z( \mathbf{x}_f, \mathbf{x}_n ) = h_f(\mathbf{x}_f) + h_n(\mathbf{x}_n)
\end{align}
, where $h_f(\cdot)$ and $h_n(\cdot)$ represent the projection functions that map respective inputs to the representation vectors and can be implemented, for instance, using dense layers.


\textbf{Attentive Representation.} 
Extending the Representation Summation method, we introduce modality-wise weights to adapt the fusion behavior across instances~\cite{nagrani2021attention, koval2024financial}, as defined in Eq.~\ref{eq:rep_weights}:
\begin{align}
& z( \mathbf{x}_f, \mathbf{x}_n ) = a_f h_f(\mathbf{x}_f) + a_n h_n(\mathbf{x}_n)  
\label{eq:rep_weights}
\\
& [a_f, a_n] 
=
\mathrm{softmax}
\big(
w( \mathbf{x}_f, \mathbf{x}_n )
\big)
\label{eq:rep_logits}
\end{align}
, where $a_f$ and $a_n$ are scalar weights, satisfying $0 \leq a_f, a_n \leq 1 $.
$h_f(\cdot)$ and $h_n(\cdot)$ are as defined in Eq.~\ref{eq:rep_sum}.
Then, in Eq.~\ref{eq:rep_logits}, $w(\cdot, \cdot)$ is a logits function producing two unnormalized scores.

\begin{wrapfigure}[15]{r}{0.66\textwidth}
  \begin{center}
    \includegraphics[width=0.66\textwidth]{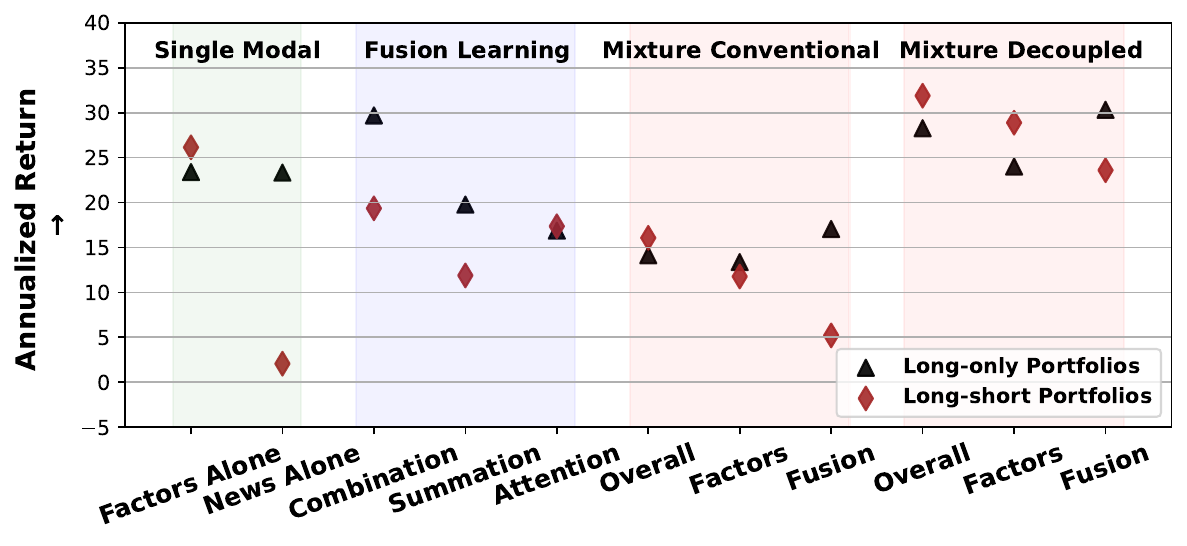}
  \end{center}
  \caption{
  Illustrative comparison of different methods' portfolio performance (North American Universe).
  }
  \label{fig:na_obs_perf}
\end{wrapfigure}

\textbf{Empirical Observations Motivating the Mixture Model.} 
We briefly discuss the illustrative comparison in Fig.~\ref{fig:na_obs_perf} and leave the full results and analysis to the experiment and appendix sections.
In Fig.~\ref{fig:na_obs_perf}, there are four blocks, each with different methods indicated on the x-axis.
In the leftmost block of Fig.~\ref{fig:na_obs_perf}, \emph{Factors Alone} and \textit{News Alone} are single-modal methods (with details in the experiment and Appendix sections).
In the Fusion Learning block, \textit{Combination}, \textit{Summation}, and \textit{Attention} correspond to the three methods presented above.
The Mixture Conventional and Mixture Decoupled blocks denote mixture models trained under different schemes, as discussed in the next subsection.

In the Fusion Learning block of Fig.~\ref{fig:na_obs_perf}, \textit{Combination}, i.e., the representation combination method, achieves superior performance compared with \textit{News Alone} in the Single Modal block.
This suggests that fusion learning can generate predictive representations, with the effectiveness depending on specific methods.
In this example, the relatively simple \textit{Combination} method outperforms complex alternatives.

However, a comparison with \textit{Factors Alone} in Fig.~\ref{fig:na_obs_perf} reveals a limitation of fusion learning.
While the \textit{Combination} method improves the long-only portfolio, it underperforms in the long-short portfolio, indicating its weaker predictive performance for low-return stocks, compared to \textit{Factors Alone}.
This behavior may arise when newsflow is less relevant or offers little incremental information for certain instances or market regimes; in such cases, fusion learning dilutes information from factors and reduces performance, as illustrated here for low-return stocks.
More generally, this dilution effect may manifest across stocks and time, as the relative relevance of factors and news evolves.


\subsubsection{Mixture Modeling over Factors-based and Fusion-based Predictions}
Based on the above analysis of Fig.~\ref{fig:na_obs_perf}, it is desirable to include factors in a separate prediction component and to adaptively leverage the factors-based and fusion-based predictions when they excel under different conditions. 
Accordingly, our mixture model comprises two prediction components, as formulated in Eqs.~\ref {eq:mixture_pred} and~\ref{eq:mixture_prob} (with implementation details provided in the Appendix).
Theoretically, a two-component mixture model can achieve prediction errors lower than those of either component alone (see Proposition 2 for details).
\begin{align}
& \hat{r} 
= 
\sum_{i \in \{f, u\} } p_{\phi}(I {=} i \,| \mathbf{x}_f, \mathbf{x}_n ) \cdot g_{\theta_i}( \mathbf{x}_i ) 
\label{eq:mixture_pred} 
\\
& \big[ p_{\phi}(I {=} f | \mathbf{x}_{f}, \mathbf{x}_n ), \, p_{\phi}(I {=} u | \mathbf{x}_{f}, \mathbf{x}_n ) \big]
= 
\mathrm{softmax}
\big(
\ell_{\phi}( \mathbf{x}_{f}, \mathbf{x}_n )
\big)
\label{eq:mixture_prob} 
\end{align}
In Eq.~\ref{eq:mixture_pred}, let $i \in \{f, u\}$ index the prediction components. 
$i = f$ refers to the factors-based prediction component $g_{\theta_f}( \mathbf{x}_f )$, and $i = u$ refers to the fusion-based prediction component $g_{\theta_u}( \mathbf{x}_u )$. 
The factors-based prediction function $g_{\theta_f}( \mathbf{x}_f )$ is implemented as a dense network parameterized by $\theta_f$.
Motivated by the competitive performance of the representation combination method in experiments, the fusion-based component $g_{\theta_u}( \cdot )$ with input $\mathbf{x}_u$ follows the formulation in Eq.~\ref{eq:fusion_framework} and Eq.~\ref{eq:rep_combine}.

The mixture weights are defined as a probability distribution $p_{\phi}(I \,| \mathbf{x}_f, \mathbf{x}_n)$ over the component index $I \in \{ f, u \}$, which facilitates the formulation of the subsequent decoupled training.
In Eq.~\ref{eq:mixture_prob}, let
$
\boldsymbol{\ell}_{\phi} : \mathbb{R}^{d_f} \times \mathbb{R}^{d_n} \to \mathbb{R}^2
$
be a logits function parameterized by $\phi$, which takes the factors $\mathbf{x}_f \in \mathbb{R}^{d_f}$ and the newsflow representation $\mathbf{x}_n \in \mathbb{R}^{d_n}$ and outputs a vector of two unnormalized scores (logits) corresponding to the two predictions.

\begin{figure}[tbp]
  \centering
  \begin{subfigure}[b]{0.655\textwidth}
    \includegraphics[width=\textwidth]{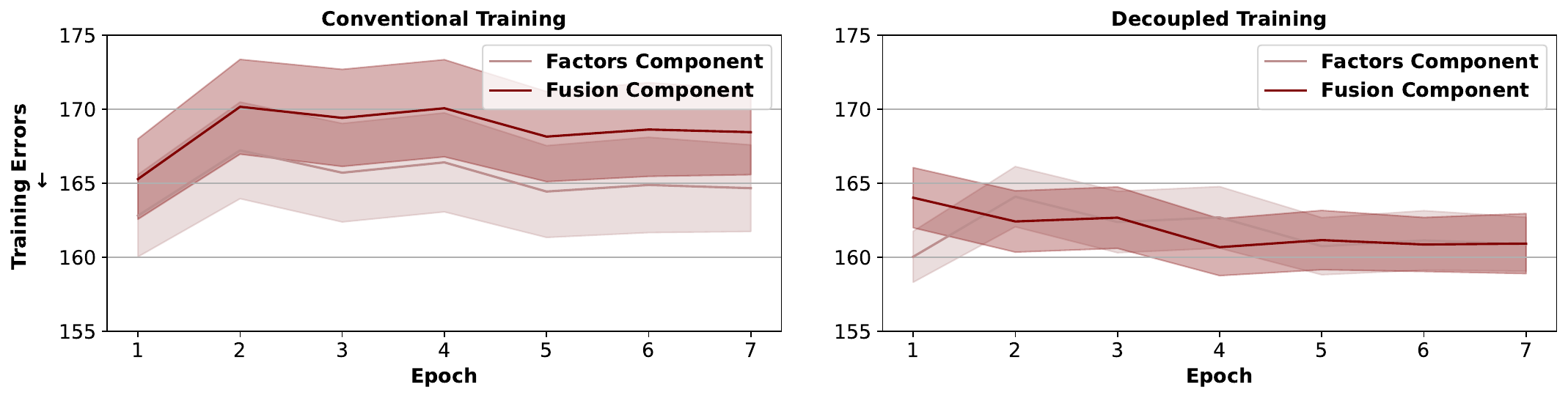}
    \caption{}
    \label{fig:na_obs_train}
  \end{subfigure}
  \hspace{0.em}
  \begin{subfigure}[b]{0.325\textwidth}
    \includegraphics[width=\textwidth]{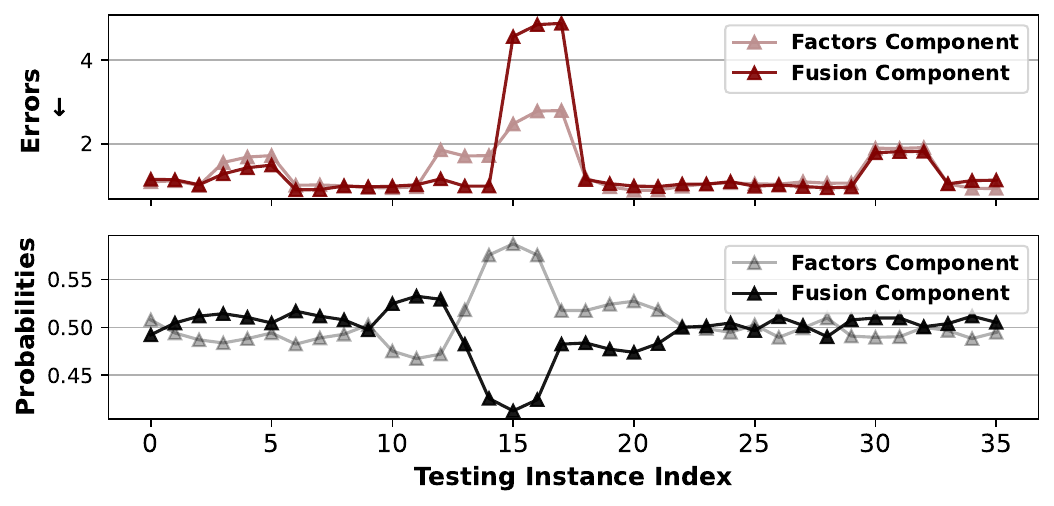}
    \caption{}
    \label{fig:na_mix_weight}
  \end{subfigure}
  
  \caption{
  (a) Training error curves of the mixture model's prediction components by different training methods.
  (b) Illustration of the alignment between each component’s prediction errors (top panel) and mixture probabilities (bottom panel) learned via the distribution matching in decoupled training.
  Note that the x-axis represents test samples ordered arbitrarily (not a time series).
  }
  \label{fig:main}
\end{figure}

\textbf{Limitations of Conventional Training.}
The conventional training of the mixture model minimizes the squared errors over training data as:
\begin{align}\label{eq:conventional_mix_loss}
\begin{split}
& \min_{\theta_f, \theta_u, \phi} \mathbb{E}_{ \{\mathbf{x}_f, \mathbf{x}_n, r\} \sim \mathcal{D} } 
\bigg[
\Big[
r - \sum_{i \in \{f, u\} } p_{\phi}(I {=} i \,| \mathbf{x}_f, \mathbf{x}_n) \cdot g_{\theta_i}(\mathbf{x}_i)
\Big]^2
\bigg]
\end{split}
\end{align}
However, empirically, we observe that this conventional training often leads to unstable convergence of individual components and degraded performance~\cite{guo2022learning}.
For instance, in Fig.~\ref{fig:na_obs_train}, under conventional training, the training error curves of the mixture model’s two prediction components exhibit slow and unstable convergence.
In the Mixture Conventional block of Fig.~\ref{fig:na_obs_perf}, \textit{Factors} and \textit{Fusion} correspond to the return predictions of the respective components within the mixture model, while \textit{Overall} represents the combined mixture predictions.
Compared with \textit{Factors Alone} and \textit{Combination} under standalone training, the \textit{Factors} and \textit{Fusion} components present performance degradation.
Consequently, the mixture model fails to achieve satisfactory overall performance, as shown in the \textit{Overall} result. 
%
%
%
%

Fundamentally, as presented in Proposition~1 (with proof and extended discussion in the Appendix), the conventional training is affected by entangled gradient variance.
\begin{proposition}[\textbf{Entangled Gradient Variance}]
Consider stochastic gradient descent with instances sampled as $\{ \mathbf{x}_f, \mathbf{x}_n, r \} \sim \mathcal{D}$.
Let $\hat{r}$ denote the model's prediction.
Let $i$ index a prediction component in the mixture model,
and define the gradient signal for component $i$ as $\zeta_i:= \left( r - \hat{r} \right) \cdot \nabla_{\theta_i} g_{\theta_i}( \mathbf{x}_i )$, where $\theta_i$ are the parameters of the $i$-th prediction component.
Define $p_i:= p_{\phi}(I {=} i | \mathbf{x}_f, \mathbf{x}_n)$.
Under the conventional training objective given in Eq.~\ref{eq:conventional_mix_loss}, the variance of the stochastic gradient $\delta_i$ used to update $\theta_i$ is:
%
%
%
%
\begin{align}\label{eq:mixture_var}
\operatorname{Var}(\delta_i) =
4 \mathbb{E}^2[p_i] \operatorname{Var}(\zeta_i)
+ 
4 \mathbb{E}[ \,\norm{\zeta_i}^2 ] \operatorname{Var}(p_i)
\end{align}
By contrast, under the standalone training of component $i$, the gradient variance is:
\begin{align}\label{eq:standalone_var}
\operatorname{Var}(\delta_i) = 4 \operatorname{Var}(\zeta_i)
\end{align}
\end{proposition}
%
%
From standard results in stochastic optimization~\cite{bubeck2015convex, bottou2018optimization}, the convergence behavior is closely tied to the variance of stochastic gradients on training data.
The convergence rate is generally bounded by $\mathcal{O}\big( {\operatorname{Var}(\delta) / \sqrt{k}} \big)$. 
$\operatorname{Var}(\delta)$ denotes the variance of stochastic gradient $\delta$, and $k$ is the iteration step.

As shown in Eq.~\ref{eq:mixture_var}, the unstable convergence of the mixture model's components is related to the gradient variance arising from the variances of $\zeta_i$ and $p_i$ with additional entanglement through their respective expectations.
This entanglement implies that even if individual variances are moderate, their interaction can amplify the total variance.
In particular, $\mathbb{E}[\|\zeta_i\|^2]$ can be large due to the number of parameters in prediction components implemented by dense networks.
Meanwhile, since all prediction components influence the residual $r-\hat{r}$ in the mixture model, an inaccurate component can inflate the residual, thereby increasing the variance.
By contrast, the standalone training of a prediction component, such as the fusion learning in Sec.~\ref{sec:fusion_learn}, has no such entanglement in Eq.~\ref{eq:standalone_var}.


\textbf{Decoupled Training.}
Given the above observations and analysis, we propose the decoupled training method (with theoretical insights in the Appendix).

The key idea is to train each prediction component independently to realize its predictive capacity, while learning the probability distribution based on the actual relative performance of each component.
Concretely, the decoupled training minimizes the loss function comprising two parts: 
%
%
%
%
%
%
%
\begin{align}
& \min_{\theta_f, \theta_u, \phi} 
\mathbb{E}_{ \{\mathbf{x}_f, \mathbf{x}_n, r\} \sim \mathcal{D} }
\Big[
\underbrace{ L\big( \theta_f, \theta_u \,; \{ \mathbf{x}_f, \mathbf{x}_n, r \} \big) }_{\text{Independent Training}}
+
\underbrace{ L\big( \phi \,; \{ \mathbf{x}_f, \mathbf{x}_n, r \}, \hat{\theta}_f, \hat{\theta}_u \big) }_{\text{Distribution Matching}}
\Big]
\label{eq:loss_decoupled}
\end{align}
\begin{align}
& L\big( \theta_f, \theta_u \,; \{ \mathbf{x}_f, \mathbf{x}_n, r \} \big) 
:=
\sum_{i \in \{f, u\}}\big[ r - g_{\theta_i}(\mathbf{x}_i) \big]^2 
\label{eq:loss_pred}
\\
& L\big( \phi \,; \{ \mathbf{x}_f, \mathbf{x}_n, r \}, \hat{\theta}_f, \hat{\theta}_u \big) 
:=
\kl
\Big[
p_{\phi}( I \,| \mathbf{x}_{f}, \mathbf{x}_n ) 
\,\big\Vert\,
p_{ \hat{\theta}_f, \hat{\theta}_u}(I \,| \mathbf{x}_f, \mathbf{x}_n, r )
\Big]
\label{eq:loss_prob}
\end{align}
The Independent Training term $L\big( \theta_f, \theta_u \,; \cdot \big)$ involves solely the parameters $\theta_f$ and $\theta_u$ for training prediction components.
It is realized by minimizing the squared error of each component in Eq.~\ref{eq:loss_pred}.

The Distribution Matching term $L\big( \phi \,; \cdot \big)$ aligns the mixture probability $p_{\phi}(I \,| \mathbf{x}_{f}, \mathbf{x}_n )$ with a target distribution $p_{ \hat{\theta}_f, \hat{\theta}_u}(I \,| \mathbf{x}_f, \mathbf{x}_n, r )$ that reflects the actual relative prediction performance of each component on each data instance.
Here, $\hat{\theta}_f$ and $\hat{\theta}_u$ denote the given parameter values of the prediction components.
In Eq.~\ref{eq:loss_prob}, minimizing the Kullback–Leibler (KL) divergence, with respect to $\phi$, encourages $p_{\phi}(I \,| \mathbf{x}_{f}, \mathbf{x}_n )$ to allocate probabilities in accordance with the relative prediction performance indicated in $p_{ \hat{\theta}_f, \hat{\theta}_u}(I \,| \mathbf{x}_f, \mathbf{x}_n, r )$.
The KL divergence between discrete distributions is analytically tractable and fits into stochastic gradient descent-based optimization~\cite{blei2017variational}.

Specifically, the target distribution $p_{ \hat{\theta}_f, \hat{\theta}_u}(I \,| \mathbf{x}_f, \mathbf{x}_n, r )$ is defined in Eq.~\ref{eq:target_distribution}:
\begin{align}\label{eq:target_distribution}
\big[
p_{\hat{\theta}_f, \hat{\theta}_u}(I {=} f | \cdot )
, \,
p_{\hat{\theta}_f, \hat{\theta}_u}(I {=} u | \cdot )
\big]
= 
\text{\small softmax}
\Big(
-\big( r-g_{\hat{\theta}_f}\big( \mathbf{x}_f ) \big)^2/\tau  \,,  -\big( r-g_{\hat{\theta}_u}( \mathbf{x}_u ) \big)^2/\tau
\Big)
\end{align}
In Eq.~\ref{eq:target_distribution}, the softmax function takes as input the prediction errors of the two components, negatively scaled by the temperature parameter $\tau$.
The output probabilities $p_{\hat{\theta}_f, \hat{\theta}_u}( I \,| \cdot )$ reflect the relative prediction performance by assigning higher probabilities to components with lower prediction errors.
For brevity, we omit $(\mathbf{x}_f, \mathbf{x}_n, r)$ in the conditioning notation of $p_{\hat{\theta}_f, \hat{\theta}_u}(I \,| \cdot )$.
The terms $g_{\hat{\theta}_f}(\mathbf{x}_f)$ and $g_{\hat{\theta}_u}(\mathbf{x}_u)$ in Eq.~\ref{eq:target_distribution} represent the respective predictions of the two components.

During training, we adopt a simple estimate for $\hat{\theta}_f$ and $\hat{\theta}_u$, i.e., using the most recent values.
For instance, at step $k$, $\hat{\theta}_f$ is set to the value of $\theta_f$ from step $k-1$.
Thus, $p_{ \hat{\theta}_f, \hat{\theta}_u}(I \,| \cdot )$ reflects the latest learned predictive performance of each component and becomes increasingly reliable as training progresses.
Exploring alternative estimation methods is left for future work.

At test time, when $r$ is not available and $p_{ \hat{\theta}_f, \hat{\theta}_u}(I \,| \cdot )$ cannot be computed, $p_{\phi}(I \,| \mathbf{x}_{f}, \mathbf{x}_n )$ serves to infer the relative performance of each component and then to combine the predictions as Eq.~\ref{eq:mixture_pred}.
For instance, Fig.~\ref{fig:na_mix_weight} illustrates the alignment learned through distribution matching: in general, components with higher prediction errors receive lower probabilities, and vice versa.

Compared with conventional training, as shown in Fig.~\ref{fig:na_obs_train}, the training error curves of the two prediction components exhibit relatively stable convergence and lower errors under decoupled training. 
Accordingly, in the Mixture Decoupled block of Fig.~\ref{fig:na_obs_perf}, the \textit{Factors} and \textit{Fusion} components achieve performance comparable to \textit{Factors Alone} and \textit{Combination}, respectively.
The overall performance of the mixture model is notably improved relative to conventional training.

\textbf{Remarks.}
The target distribution defined in Eq.~\ref{eq:target_distribution} is motivated by a probabilistic inference perspective and connects to variational inference, as discussed in the Appendix~\ref{appendix:theory}. 
Alternative ways of formulating the target distribution can be based on the closed-form optimal mixture weight derived during training, which is related to Eq.~\ref{eq:target_distribution} as presented in Proposition 2 (with proof in the Appendix).
In particular, Proposition 2 shows that Eq.~\ref{eq:target_distribution} is a monotone transformation of the same underlying signal $\hat{p}$ as the optimal weight, thereby preserving the ordering of the optimal weights across mixture components.
Moreover, whereas the optimal weight can be binary, the target distribution remains relatively smooth for optimization, reducing the risk of overfitting.
We leave the comparison of different instantiations of target distributions for future work.
\begin{proposition}[\textbf{Optimal Mixture Weights and the Relation to Eq.~\eqref{eq:target_distribution}}\,]
For a data instance $(\mathbf{x}_f, \mathbf{x}_n, r)\sim\mathcal{D}$, let $\hat{r}_f$ and $\hat{r}_u$ denote the factors-based and fusion-based predictions under parameters $\hat{\theta}_f$ and $\hat{\theta}_u$.
Define
$
\hat{p}:=\frac{r - \hat{r}_u}{\hat{r}_f - \hat{r}_u}
$
and
$
d:=\hat{r}_f - \hat{r}_u
$.
Then, the optimal mixture weight minimizing squared errors is
$
p_f^* = \operatorname{clip}_{[0,1]}(\hat{p})
$
and
$
p_u^* = 1 - p_f^*
$
, and the mixture prediction
$\hat{r}^* = p_f^* \hat{r}_f + p_u^* \hat{r}_u$
satisfies
$
\mathbb{E}\!\left[(r-\hat{r}^*)^2\right]
\le
\operatorname{min}\!\left\{
\mathbb{E}\!\left[(r-\hat{r}_f)^2\right]
,
\mathbb{E}\!\left[(r-\hat{r}_u)^2\right]
\right\}
$.
The target distribution defined in Eq.~\eqref{eq:target_distribution} admits an equivalent expression with $\hat{p}$:
$
p_{\hat{\theta}_f,\hat{\theta}_u}(I \,| \mathbf{x}_f, \mathbf{x}_n, r)
=
\left[
\sigma\!\Big( \frac{d^2(2\hat{p}-1)}{\tau} \Big),
\;
1 - \sigma\!\Big( \frac{d^2(2\hat{p}-1)}{\tau} \Big)
\right]
$
, where $\sigma( a )$ is the sigmoid function $1/(1 + e^{-a})$.
\end{proposition}
%
%



\section{Experiments}\label{sec:exp}
In this section, we briefly present the experiment setup and primarily discuss the results from two investment universes.
The full experiment details and results for all universes are in the Appendix.


\textbf{Data.}
We have three datasets corresponding to the North American (NA), Emerging Markets (EM), and European (EU) investment universes, each containing up to $\sim1,000$ stocks. 
We use company-level financial news data provided by a commercial data vendor.
In the Appendix, Table~\ref{tab:factors_categories} and ~\ref{tab:news_categories} list main categories of factors and news in our data, while Table~\ref{tab:data_size} presents the statistics of training, validation, and testing data.


\textbf{Baselines.}
For a fair comparison, we employ the encoder-only LLM, DeBERTa~\cite{he2021debertav3}, across our fusion learning methods, mixture models, and all LLM-based baselines. 
\underline{\textit{Universe}} refers to a portfolio that equally weights all stocks in the investment universe. 
\underline{\textit{Factors Alone}} represents a dense neural network solely on quantitative factors~\cite{chen2024deep}.
Note that for a fair comparison, the mixture model's factor-based prediction component adopts the same model structure as this baseline.
\underline{\textit{News Alone}} utilizes only news by employing a prediction layer on the news representations generated by an LLM~\cite{guo2024fine}.
\underline{\textit{FININ}} develops a factors-based attention to weight newsflow representations before jointly passing both through a prediction layer~\cite{wang2024modeling}.


\textbf{Setup.}
For training, we use the one-month forward return as the target variable, as the subsequent backtest focuses on monthly rebalanced portfolios.
After training, the model is evaluated on the testing period without retraining in a rolling manner.
The testing period covers 2023 and 2024 for mitigating potential memorization bias in LLMs~\cite{rahimikia2024r, levy2024caution, lopez2025memorization} (with explanations in the Appendix).
For fine-tuning, we applied Low-Rank Adaptation (LoRA) to all layers of DeBERTa~\cite{hu2021lora,ding2023parameter, rasley2020deepspeed}.

During backtesting, the long-only portfolio consists of stocks in the top (9th) decile of predicted returns, while the long-short portfolio holds stocks in both the top (9th) and bottom (0th) deciles~\cite{guo2024fine, chauhan2020uncertainty}.
Both portfolios are equally weighted and rebalanced monthly.


\textbf{Metrics.}
We use annualized returns and Sharpe ratios for evaluating portfolio performance, and Mean Absolute Percentage Error (MAPE) and the Information Coefficient (IC) as prediction metrics~\cite{duan2022factorvae}.
Meanwhile, we present the bar charts of decile returns to illustrate the sources of portfolio performance.
Additionally, we report the results without and with LLM fine-tuning.


\begin{table}[t]
  \centering
  \caption{
  Portfolio and prediction performance.
  The best and second-best results are highlighted with dark gray \rankone{\,\,} and light gray \ranktwo{\,\,} boxes, respectively.
  }
  \begin{subtable}{\textwidth}
    \caption{
    North American Universe
    }
    \resizebox{1.\textwidth}{!}{
      \begin{tabular}{|c|c|c|c|c|c|c|}
        \hline
        \multirow{2}{*}{} & \multicolumn{2}{|c|}{Long-only Portfolios} & \multicolumn{2}{|c|}{Long-short Portfolios} & \multicolumn{2}{|c|}{Prediction Metrics}\\
        \cline{2-7}
        & Ann. Return \% ($\uparrow$) & Sharpe Ratio ($\uparrow$) & Ann. Return \% ($\uparrow$) & Sharpe Ratio ($\uparrow$) & MAPE ($\downarrow$) & IC ($\uparrow$) \\
        \hline
    Universe       & 12.37 & 0.84 & $-$  & $-$  & $-$ & $-$  \\
    Factors Alone & 22.31 & 0.81 & 22.34 & 1.26 & 1.352 & 0.018 \\
    News Alone & 20.96 & \rankone{1.03} & 1.08 & 0.18 & \rankone{1.092} & -0.000 \\
    FININ & 23.12 & 0.83 & 18.16 & 1.16 & 1.467 & 0.019 \\
    \hline
    Fusion Combination & \rankone{32.43} & \ranktwo{1.0} & 28.41 & \ranktwo{1.64} & 1.402 & \rankone{0.031} \\
    Fusion Summation & 20.42 & 0.76 & 16.13 & 1.03 & 1.465 & 0.017 \\
    Fusion Attention & 21.73 & 0.73 & 19.59 & 0.87 & \ranktwo{1.302} & -0.005 \\
    \hline
    Mixture Conventional & 23.27 & 0.75 & \ranktwo{29.32} & 1.42 & 1.539 & 0.016 \\
    Mixture Decoupled & \ranktwo{28.21} & 0.92 & \rankone{33.77} & \rankone{1.78} & 1.319 & \ranktwo{0.027} \\
      \hline
      \end{tabular}
    }
    \label{tab:na_portfolio}
  \end{subtable}
  %
  \begin{subtable}{\textwidth}
    \caption{
    Emerging Markets Universe
    }
    \resizebox{1.\textwidth}{!}{
      \begin{tabular}{|c|c|c|c|c|c|c|}
        \hline
        \multirow{2}{*}{} & \multicolumn{2}{|c|}{Long-only Portfolios} & \multicolumn{2}{|c|}{Long-short Portfolios} & \multicolumn{2}{|c|}{Prediction Metrics}\\
        \cline{2-7}
        & Ann. Return \% ($\uparrow$) & Sharpe Ratio ($\uparrow$) & Ann. Return \% ($\uparrow$) & Sharpe Ratio ($\uparrow$) & MAPE ($\downarrow$) & IC ($\uparrow$) \\
        \hline
    Universe       & 2.63 & 0.24 & $-$  & $-$    & $-$ & $-$  \\
    Factors Alone & \ranktwo{17.14} & \ranktwo{0.81} & \rankone{42.17} & \rankone{2.96} & 1.461 & 0.049 \\
    News Alone & -3.94 & -0.2 & -9.67 & -1.57 & \rankone{1.194} & -0.015 \\
    FININ & 12.9 & 0.72 & 30.02 & 2.15 & 1.445 & 0.046 \\
    \hline
    Fusion Combination & 13.36 & 0.75 & 32.35 & 2.21 & 1.452 & \ranktwo{0.060} \\
    Fusion Summation & 13.38 & 0.74 & 28.52 & 1.96 & 1.474 & 0.043 \\
    Fusion Attention & 11.85 & 0.67 & 14.22 & 1.07 & \ranktwo{1.283} & 0.003 \\
    \hline
    Mixture Conventional & 7.71 & 0.46 & 22.47 & 1.53 & 1.465 & 0.053 \\
    Mixture Decoupled & \rankone{18.5} & \rankone{0.94} & \ranktwo{42.07} & \ranktwo{2.93} & 1.395 & \rankone{0.065} \\
      \hline
      \end{tabular}
    }
    \label{tab:em_portfolio}
  \end{subtable}
  \label{tab:portfolio}
\end{table}

\begin{figure}[tp]
  \centering
  \begin{subfigure}[b]{0.49\textwidth}
    \includegraphics[width=\textwidth]{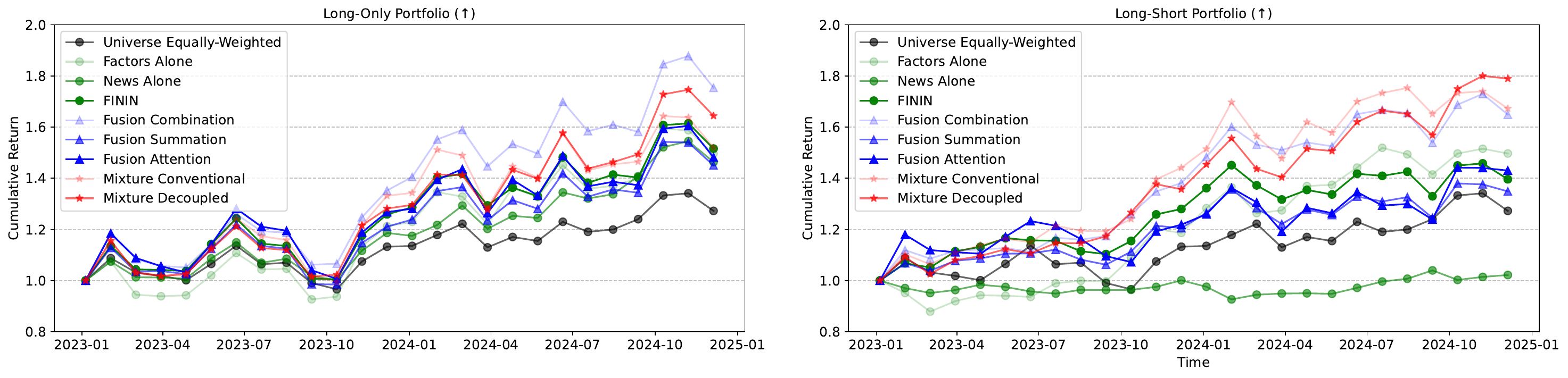}
    \caption{North American Universe}
    \label{fig:na_deciles}
  \end{subfigure}
  \begin{subfigure}[b]{0.49\textwidth}
    \includegraphics[width=\textwidth]{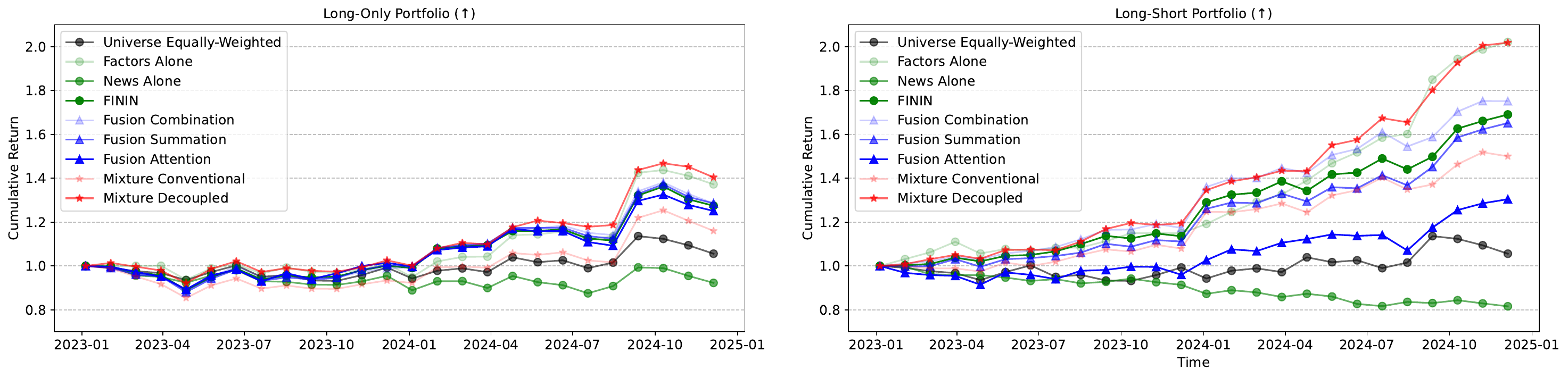}
    \caption{Emerging Markets Universe}
    \label{fig:em_deciles}
  \end{subfigure}
  \caption{Performance Charts.
  }
  \label{fig:charts}
\end{figure}

\textbf{Results.} 
Table~\ref{tab:portfolio} reports the portfolio and prediction performance, while Fig.~\ref{fig:charts} visualizes the cumulative returns of corresponding portfolios. 
\textit{Fusion Combination}, \textit{Fusion Summation}, and \textit{Fusion Attention} refer to the three methods in fusion learning.
\textit{Mixture Conventional} and \textit{Mixture Decoupled} represent the mixture models trained under the conventional and decoupled training schemes.

\underline{Portfolio Performance of Fusion Learning.} 
Within the Fusion group of Table~\ref{tab:portfolio}, \textit{Fusion Combination} achieves superior performance in long-only and long-short portfolios.
The weaker portfolio performance of other fusion methods may stem from their tendency to compress heterogeneous modalities into a shared representation space, potentially obscuring or destroying complementary information~\cite{chaudhuricloser}.
The precise cause necessitates investigation in future work.
These results suggest a simple yet effective principle for designing fusion methods: preserving the structure of each modality and learning across the set of modality-specific representations can yield predictive representations.

Furthermore, comparing \textit{Fusion Combination} with \textit{News Alone} and \textit{Factors Alone} across Tables~\ref {tab:na_portfolio} and~\ref{tab:em_portfolio} reveals that the effectiveness of fusion learning is universe-dependent, influenced by the distinct relative relevance of factors and news across markets.
This observation echoes the motivation of the mixture model, whose results are discussed below.

Specifically, \textit{Fusion Combination} consistently outperforms \textit{News Alone}, reflecting the notable predictive power introduced by factors.
However, while \textit{Fusion Combination} performs competitively compared with \textit{Factors Alone} in the NA universe (Tables~\ref {tab:na_portfolio}), it lags in the EM universe (Tables~\ref {tab:em_portfolio}).
This implies that in the NA universe, news data provides complementary information to factors, leading to improved performance of \textit{Fusion Combination}.
In contrast, in the EM universe, our news data appears to provide little incremental information on factors. 
In such cases, as discussed in Fig.~\ref{fig:na_obs_perf}, fusion learning struggles to fully leverage the predictive power of factors through the unified representations, resulting in underperformance.


\underline{Portfolio Performance of Mixture Models.} 
In the Mixture group of Table~\ref{tab:na_portfolio}, \textit{Mixture Decoupled} demonstrates competitive performance and robustness across portfolios, indicating that decoupled training helps unlock the potential of the mixture model.
Specifically, in long-only portfolios, \textit{Mixture Decoupled} trails the top-performing \textit{Fusion Combination}.
In long-short portfolios, \textit{Mixture Decoupled} becomes the best-performing method and improves upon its long-only results, indicating that the short part of the portfolio contributes positively. 

In Table~\ref{tab:em_portfolio}, in contrast to the underperformance of \textit{Fusion Combination} relative to \textit{Factors Alone}, \textit{Mixture Decoupled} ranks highest in the long-only portfolios and marginally trails \textit{Factors Alone} in the long-short portfolios.
By disentangling the predictions from factors and fusion and adaptively combining them for individual data instances, \textit{Mixture Decoupled} becomes less sensitive to varying data characteristics across universes, thereby retaining the competitive performance


%
%
\begin{figure}[!htbp]
  \centering
  \begin{subfigure}[b]{0.45\textwidth}
    \includegraphics[width=\textwidth]{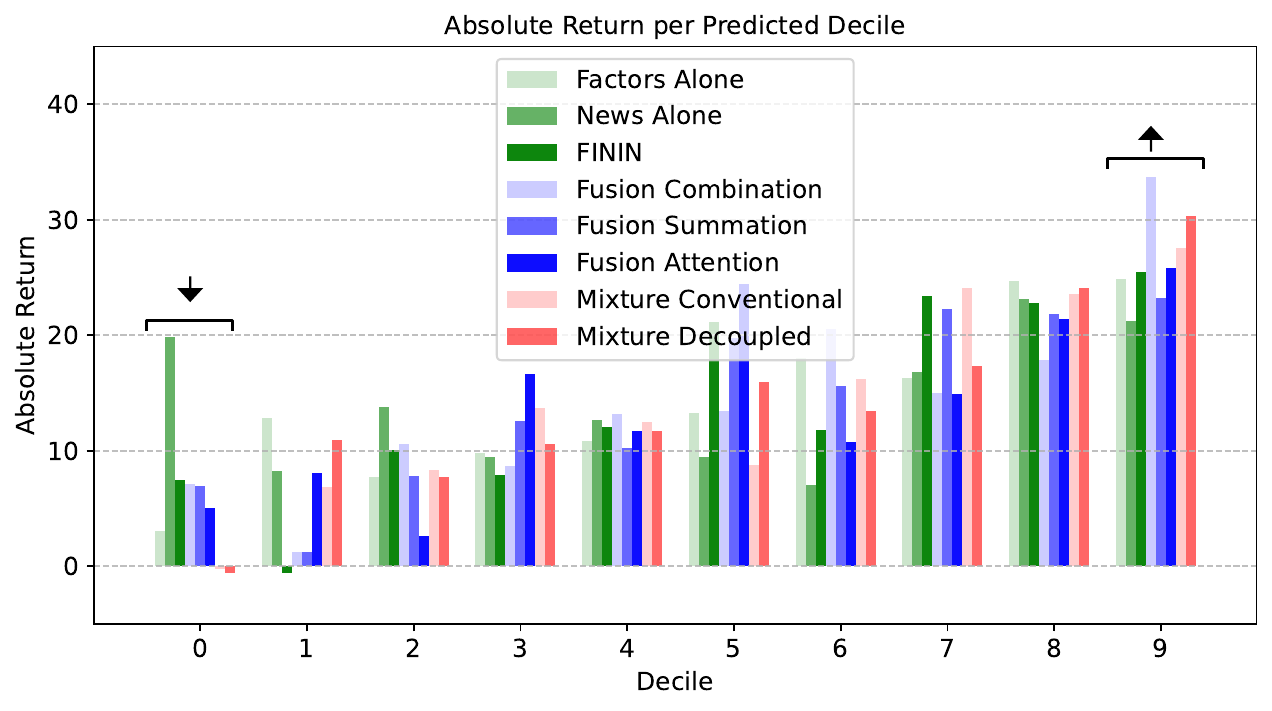}
    \caption{North American Universe}
    \label{fig:na_deciles}
  \end{subfigure}
  \begin{subfigure}[b]{0.45\textwidth}
    \includegraphics[width=\textwidth]{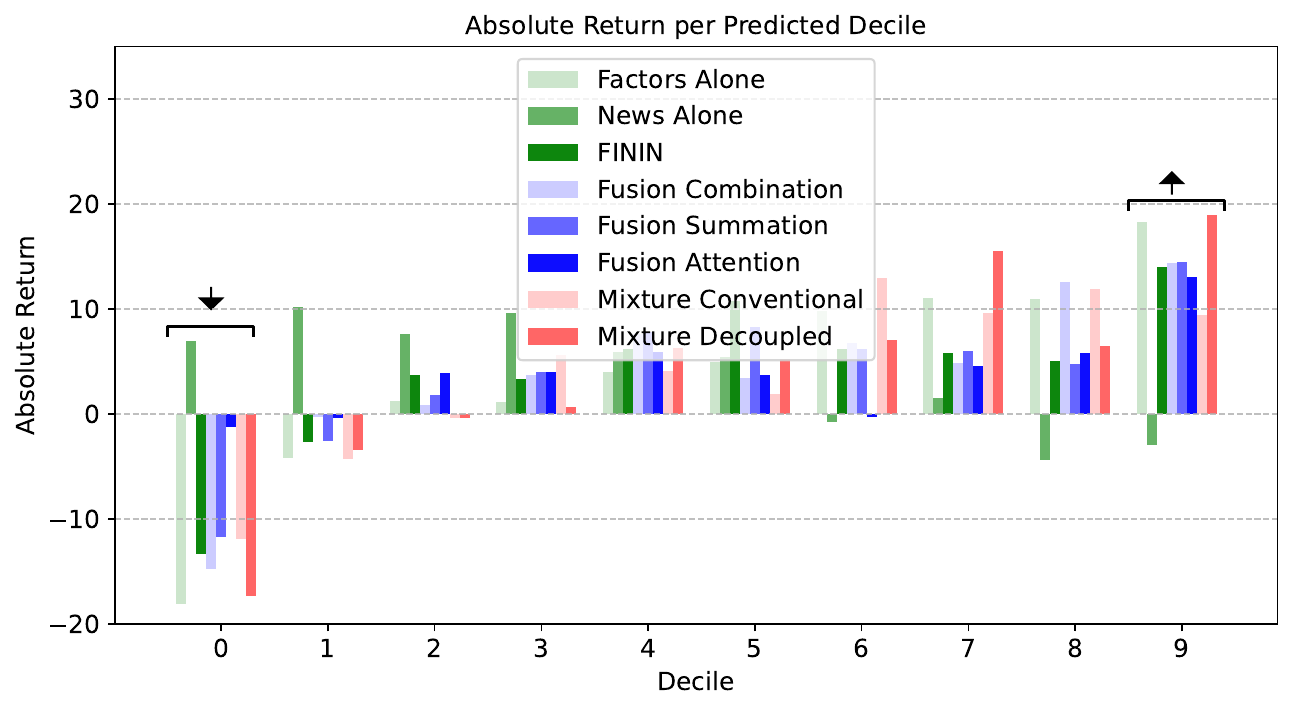}
    \caption{Emerging Markets Universe}
    \label{fig:em_deciles}
  \end{subfigure}
  \caption{
  Decile Returns.
  The arrows on the 0th and 9th deciles indicate the desired value direction.
  A lower return is preferred for the 0th decile, as it represents the short leg of long-short portfolios.
  }
  \label{fig:deciles}
\end{figure}
\underline{Gap between Prediction Errors and Stock Selection Efficacy.}
The MAPE results in Table~\ref{tab:portfolio} demonstrate that a low MAPE does not guarantee a high-performing portfolio, highlighting the difference between a model's predictive error and its practical effectiveness in stock selection~\cite{dessain2022machine, hubavcek2023beating}.
For instance, consider the \textit{Fusion Attention} method in Table~\ref{tab:na_portfolio}, which has a relatively low MAPE; yet it delivers a weaker portfolio performance.

MAPE assesses average prediction errors across test data and is largely insensitive to the relative ordering of predicted returns, which is crucial for stock selection.
Moreover, MAPE is symmetric: it penalizes over- and under-predictions of the same magnitude equally.
For instance, in stock selection, a small under-prediction may push a truly high-return stock out of the top decile. 
At the same time, a comparable over-prediction may incorrectly include a mid-return stock into the top decile.
Thus, a model with a low MAPE may still fail to identify the high-return or low-return stocks, leading to suboptimal portfolio performance.


\underline{IC as a Relevant Indicator.}
The IC results in Table~\ref{tab:portfolio} provide an indicator relevant to portfolio performance~\cite{zhang2020information, duan2022factorvae}.
Although these IC values may appear small, in quantitative finance, a small positive IC can indicate meaningful predictive power and lead to effective stock selection on a large universe of stocks.
For instance, in Table~\ref{tab:na_portfolio}, \textit{Fusion Combination} achieves the highest IC, followed by \textit{Mixture Decoupled}, consistent with their strong performance in long-only and long-short portfolios.


\begin{table}[!t]
  \centering
  \caption{
  Portfolio and prediction performance without and with enabling LLM fine-tuning during training.
  The best and second-best results within each group (without fine-tuning and with fine-tuning) are highlighted with dark gray \rankone{\,\,} and light gray \ranktwo{\,\,} boxes.
  }
  \begin{subtable}{\textwidth}
    \caption{
    North American Universe
    }
    \resizebox{1.\textwidth}{!}{
  \begin{tabular}{|c|c|c|c|c|c|c|c|}
    \hline
    \multirow{2}{*}{} &  & \multicolumn{2}{|c|}{Long-only Portfolios} & \multicolumn{2}{|c|}{Long-short Portfolios} & \multicolumn{2}{|c|}{Prediction Metrics}\\
    \cline{3-8}
    & & Ann. Return \% ($\uparrow$) & Sharpe Ratio ($\uparrow$) & Ann. Return \% ($\uparrow$) & Sharpe Ratio ($\uparrow$) & MAPE ($\downarrow$) & IC ($\uparrow$) \\
    \hline
    \multirow{7}{*}{\rotatebox{90}{w/o Fine-tuning}}
    &News Alone & 20.96 & \rankone{1.03} & 1.08 & 0.18 & \rankone{1.092} & -0.0 \\
    &FININ & 23.12 & 0.83 & 18.16 & 1.16 & 1.467 & 0.019 \\
    &Fusion Combination & \rankone{32.43} & \ranktwo{1.0} & 28.41 & \ranktwo{1.64} & 1.402 & \rankone{0.031} \\
    &Fusion Summation & 20.42 & 0.76 & 16.13 & 1.03 & 1.465 & 0.017 \\
    &Fusion Attention & 21.73 & 0.73 & 19.59 & 0.87 & \ranktwo{1.302} & -0.005 \\
    &Mixture Conventional & 23.27 & 0.75 & \ranktwo{29.32} & 1.42 & 1.539 & 0.016 \\
    &Mixture Decoupled & \ranktwo{28.21} & 0.92 & \rankone{33.77} & \rankone{1.78} & 1.319 & \ranktwo{0.027} \\
    \hline
    \multirow{7}{*}{\rotatebox{90}{w/ Fine-tuning}}
    &News Alone & \ranktwo{27.98} & \rankone{1.33} & 14.23 & 1.52 & \rankone{1.118} & 0.018 \\
    &FININ & 22.09 & 0.81 & 19.65 & 1.24 & 1.468 & 0.02 \\
    &Fusion Combination & \rankone{28.61} & \ranktwo{0.95} & 26.82 & \ranktwo{1.63} & 1.379 & \rankone{0.032} \\
    &Fusion Summation & 20.69 & 0.77 & 18.31 & 1.12 & 1.477 & 0.017 \\
    &Fusion Attention & 17.12 & 0.63 & 13.55 & 0.7 & \ranktwo{1.298} & -0.004 \\
    &Mixture Conventional & 26.57 & 0.85 & \rankone{32.23} & 1.43 & 1.326 & 0.017 \\
    &Mixture Decoupled & 27.15 & 0.91 & \ranktwo{30.66} & \rankone{1.79} & 1.336 & \ranktwo{0.028} \\
    \hline
  \end{tabular}
  }
  \label{tab:na_finetune}
  \end{subtable}
  \begin{subtable}{\textwidth}
    \caption{
    Emerging Markets Universe
    }
    \resizebox{1.\textwidth}{!}{
  \begin{tabular}{|c|c|c|c|c|c|c|c|}
    \hline
    \multirow{2}{*}{} &  & \multicolumn{2}{|c|}{Long-only Portfolios} & \multicolumn{2}{|c|}{Long-short Portfolios} & \multicolumn{2}{|c|}{Prediction Metrics}\\
    \cline{3-8}
    & & Ann. Return \% ($\uparrow$) & Sharpe Ratio ($\uparrow$) & Ann. Return \% ($\uparrow$) & Sharpe Ratio ($\uparrow$) & MAPE ($\downarrow$) & IC ($\uparrow$) \\
    \hline
    \multirow{7}{*}{\rotatebox{90}{w/o Fine-tuning}}
    &News Alone & -3.94 & -0.2 & -9.67 & -1.57 & \rankone{1.194} & -0.015 \\
    &FININ & 12.9 & 0.72 & 30.02 & 2.15 & 1.445 & 0.046 \\
    &Fusion Combination & 13.36 & \ranktwo{0.75} & \ranktwo{32.35} & \ranktwo{2.21} & 1.452 & \ranktwo{0.06} \\
    &Fusion Summation & \ranktwo{13.38} & 0.74 & 28.52 & 1.96 & 1.474 & 0.043 \\
    &Fusion Attention & 11.85 & 0.67 & 14.22 & 1.07 & \ranktwo{1.283} & 0.003 \\
    &Mixture Conventional & 7.71 & 0.46 & 22.47 & 1.53 & 1.465 & 0.053 \\
    &Mixture Decoupled & \rankone{18.5} & \rankone{0.94} & \rankone{42.07} & \rankone{2.93} & 1.395 & \rankone{0.065} \\
    \hline
    \multirow{7}{*}{\rotatebox{90}{w/ Fine-tuning}}
    &News Alone & 0.38 & 0.1 & 6.43 & 0.87 & \rankone{1.13} & 0.02 \\
    &FININ & 13.4 & 0.76 & \ranktwo{31.53} & \ranktwo{2.3} & 1.448 & 0.048 \\
    &Fusion Combination & 12.89 & 0.74 & 30.18 & 2.0 & 1.466 & \ranktwo{0.06} \\
    &Fusion Summation & \ranktwo{14.87} & \ranktwo{0.81} & \ranktwo{31.53} & 2.28 & 1.49 & 0.045 \\
    &Fusion Attention & 12.36 & 0.7 & 15.45 & 1.22 & \ranktwo{1.266} & 0.003 \\
    &Mixture Conventional & 13.03 & 0.73 & 27.18 & 1.84 & 1.484 & 0.049 \\
    &Mixture Decoupled & \rankone{18.18} & \rankone{0.93} & \rankone{43.49} & \rankone{3.03} & 1.403 & \rankone{0.065} \\
    \hline
  \end{tabular}
  }
  \label{tab:em_finetune}
  \end{subtable}
  
  \label{tab:finetune}
\end{table}
\underline{Decile-Level Comparison.} 
Fig.~\ref{fig:deciles}, bar charts of decile returns, provide a granular view of the investment performance across the deciles of predicted returns~\cite{guo2024fine} and illustrate the sources of the portfolio performance in Table~\ref{tab:portfolio}.
Specifically, a decile return, or the average return per predicted decile, is obtained by sorting stocks based on their predicted returns and then grouping them into ten deciles, labeled $0$ through $9$.
The 0th decile contains the stocks with the lowest predicted returns, while the 9th decile includes those with the highest predicted returns.
For each decile, we then compute the average return of the stocks within that group.

Ideally, the decile returns should exhibit a strong spread: very negative (or low) returns for the 0th decile (the short leg) and very high returns for the 9th decile (the long leg).
For instance, in Fig.~\ref{fig:na_deciles}, \textit{Fusion Combination} and \textit{Mixture Decoupled} achieve high returns in the 9th decile for the long leg, consistent with their long-only portfolio performance in Table~\ref{tab:na_portfolio}.
The distinction lies in the short leg of the portfolio: only \textit{Mixture Decoupled} delivers the desired negative return in the 0th decile, whereas \textit{Fusion Combination} still produces a positive return.
This explains why \textit{Mixture Decoupled} outperforms \textit{Fusion Combination} in the long-short portfolio: the short leg contributes more effectively, enhancing the overall long-short spread.


\underline{Inconsistent Impact of Fine-Tuning the LLM.} 
Table~\ref{tab:finetune} reports the portfolio and prediction performance without and with enabling LLM fine-tuning during the training of prediction models.
Only the methods using an LLM are included in Table~\ref{tab:finetune}, and the results without fine-tuning are from Table~\ref{tab:portfolio}.
Corresponding results for the EU universe are in Tables~\ref{tab:appendix_eu_finetune} in the Appendix.

The results in Table~\ref{tab:na_finetune},~\ref{tab:em_finetune}, and~\ref{tab:appendix_eu_finetune} demonstrate that enabling fine-tuning during training has a universe-dependent impact~\cite{brief2024mixing}.
The NA universe is a highly efficient market where public information, like news, is often quickly priced in and absorbed into factors~\cite{alves2020collective}. 
In this context, fine-tuning the LLM in multimodal models might cause it to overemphasize the already-priced or noisy news information, thereby marginally affecting or even weakening performance.
In contrast, EM and EU universes tend to be more heterogeneous and less efficient~\cite{calomiris2019news, fang2024cross}.
In these markets, news may contain more nuanced and unpriced information, making fine-tuning act as a specialization process with the potential for improvement.
These findings highlight the need for future research into adaptive fine-tuning strategies tailored to the different characteristics of investment universes.

For instance, in Table~\ref{tab:na_finetune}, for most of the multimodal methods (Fusion and Mixture groups), fine-tuning appears to have a detrimental effect.
Conversely, the single-modal \textit{News Alone} method, which relies solely on the LLM without factors, shows a notable improvement.
With fine-tuning enabled, \textit{Fusion Combination} remains the top performer in the long-only portfolio, although its performance declines.
Within the Mixture group, \textit{Mixture Decoupled} experiences a more pronounced performance drop, although its Sharpe ratios remain strong relative to \textit{Mixture Conventional}.
This may be because, in the \textit{Mixture Decoupled}, the fusion-based component is obtained through the independent training term in Eq.~\ref{eq:loss_decoupled} and is therefore more susceptible to the adverse effects of fine-tuning, consistent with the observations in the Fusion group.


\section{Conclusion}
This paper explores effective model designs and training schemes for utilizing multimodal factors and newsflow for return prediction and stock selection.
First, we introduce a representation-level fusion learning framework, realized through three representative methods.
Second, given the limitation of fusion learning observed in empirical comparison, we propose a mixture model with a specialized training scheme to mitigate the training instability.
Experiments across different investment universes reveal the following \textbf{findings}:

(1)
The competitive performance of fusion learning confirms that integrating factors and news yields predictive representations, though the choice of fusion methods is crucial.
In our experiments, the representation combination method, despite its relatively simple architecture, generally outperforms complex alternatives.
This suggests that, in noisy financial environments, effective fusion can be achieved by using simple neural networks that operate across the set of modality-specific representations.
Meanwhile, the relative performance of fusion methods varies across universes, reflecting differences in the predictive relevance of news data across markets.

(2)
The mixture model exhibits relatively robustness and balance across universes and portfolios, delivering comparable or superior performance.
Its adaptive nature can be advantageous in universes or settings where the relative predictive relevance of factors and news tends to be more variable.
Moreover, the performance improvement from the decoupled training highlights the importance of specialized training schemes for models involving entangled components.

(3)
Contrary to the intuition that fine-tuning an LLM would consistently enhance performance, our results reveal an inconsistent impact of fine-tuning during the training of our multimodal models.
This appears to depend on market efficiency and data characteristics.
In highly efficient markets like the NA universe, fine-tuning tends to overfit news information that is likely already priced in by factors, thereby degrading overall performance.
In contrast, in heterogeneous and less efficient markets, such as the EM and EU universes, fine-tuning yields performance improvements.


\newpage
\section{Future Work}
Open problems remain for future research, for instance: 
\begin{itemize}
\item Financial text often contains less relevant information that can lead models to capture spurious relations. 
Given the instruction-following capabilities of LLMs~\cite{xie2023pixiu}, it is promising to explore their use for improving data quality, for instance, by filtering, cleaning, or summarizing news content.
Such preprocessing can help enhance downstream prediction tasks and potentially improve the risk profile of portfolios constructed based on return predictions.

\item Several recent LLMs for text embedding have demonstrated strong performance~\cite{sturua2024jina, qwen3embedding}. 
It would be valuable to compare the proposed methods in this paper with these new LLMs.

\item While the mixture model adaptively combines predictions, further work could exploit diverse contextual data, such as market regimes and macroeconomic environments~\cite{antulov2024dynamic}, to inform and refine the probability weighting and distribution matching.

\item It would also be worthwhile to evaluate how the proposed methods perform when additional data sources are incorporated, such as earnings call transcripts, market time series, and so on~\cite{koval2024financial, zhang2025camef}.

\item The inconsistent impact of fine-tuning the LLM during the training of multimodal prediction models highlights the need for future research into adaptive fine-tuning strategies tailored to the characteristics of different investment universes~\cite{ke2025demystifying}.
\end{itemize}


\bibliographystyle{plain}
\bibliography{reference}

\onecolumn
\newpage
\appendix
\addcontentsline{toc}{section}{Appendix} 
\part{Appendix} 
\parttoc 


\section{Proof of Proposition 1 and Discussion}

\textbf{Proposition 1. (Entangled Gradient Variance)}
\textit{
Consider stochastic gradient descent with data instances sampled as $\{ \mathbf{x}_f, \mathbf{x}_n, r \} \sim \mathcal{D}$.
Let $\hat{r}$ denote the model's prediction.
Let $i$ index a prediction component in the mixture model,
and define the gradient signal for component $i$ as $\zeta_i:= \left( r - \hat{r} \right) \cdot \nabla_{\theta_i} g_{\theta_i}( \mathbf{x}_i )$, where $\theta_i$ are the parameters of the $i$-th prediction component.
Define $p_i:= p_{\phi}(I = i | \mathbf{x}_f, \mathbf{x}_n)$.
Under the conventional training objective given in Eq.~\ref{eq:conventional_mix_loss}, the variance of the stochastic gradient $\delta_i$ used to update $\theta_i$ is:
\[
\operatorname{Var}(\delta_i) =
4 \mathbb{E}^2[p_i] \operatorname{Var}(\zeta_i)
+ 
4 \mathbb{E}[ \,\norm{\zeta_i}^2 ] \operatorname{Var}(p_i)
\]
By contrast, under the standalone training of component $i$, the gradient variance is:
\[
\operatorname{Var}(\delta_i) = 4 \operatorname{Var}(\zeta_i)
\]
}

\begin{proof}
When training the mixture model using stochastic gradient descent-based optimization, given one training instance $\{ \mathbf{x}_f, \mathbf{x}_n, r \}$, the prediction is 
$\hat{r} = \sum_{i \in \{f,c\}} p_\phi(I=i \mid \mathbf{x}_f, \mathbf{x}_n) \cdot g_{\theta_i}( \mathbf{x}_i )$, and the joint squared loss is $ [ r - \hat{r} ]^2 $.

The stochastic gradient of the squared loss with respect to the prediction component $i$ is 
\begin{align}
  \delta_i = -2 p_i (r - \hat{r}) \, \nabla_{\theta_{i}} g_{\theta_i}( \mathbf{x}_i )
\end{align}
where $p_i := p_\phi(I = i \mid \mathbf{x}_f, \mathbf{x}_n)$.

Let $\zeta_i$ be the gradient multiplied by the prediction residual, i.e., $\zeta_i:= ( r - \hat{r} ) \nabla_{\theta_i} \hat{r} = ( r - \hat{r} ) \nabla_{\theta_i} g_{\theta_i}(\mathbf{x}_i)$, and thus the gradient of the prediction component's parameters $\theta_i$ takes the form:
\begin{equation}
    \delta_i = -2 p_i \cdot \zeta_i
\end{equation}

Note that $\delta_i$ is a vector.
For a vector $\mathbf{a} \in \mathbb{R}^{d}$, we calculate the variance of $\mathbf{a}$ as the expected squared norm of the difference between $\mathbf{a}$  and its expectation as:
$$
\text{Var}( \mathbf{a} ) 
= \mathbb{E}\Big[
\bignorm{ \mathbf{a} - \mathbb{E}[\mathbf{a}] }^2
\Big]
$$
Meanwhile, we have the identity for the variance of the product of two independent variables as:
$$
\operatorname{Var}[XY] = \operatorname{Var}[X]\operatorname{Var}[Y]
+\operatorname{Var}[X] \mathbb{E}^2[Y]
+\operatorname{Var}[Y] \mathbb{E}^2[X]
$$

Therefore, the variance of the stochastic gradient $\delta_i$ over the training data distribution is:
\begin{align}
\operatorname{Var}(\delta_i) 
& = 4 \Big[
    \mathbb{E}^2[p_i] \operatorname{Var}(\zeta_i) 
    + \bignorm{\mathbb{E}[\zeta_i]}^2 \operatorname{Var}(p_i) 
    + \operatorname{Var}(p_i) \operatorname{Var}(\zeta_i)
\Big] 
\\
& = 4 \big[ \mathbb{E}^2[p_i] + \operatorname{Var}(p_i) \big] \operatorname{Var}(\zeta_i)
+ 
4 \bignorm{\mathbb{E}[\zeta_i]}^2 \operatorname{Var}(p_i)
\\
& = 4 \mathbb{E}^2[p_i] \operatorname{Var}(\zeta_i)
+ 
4 \mathbb{E}[ \norm{\zeta_i}^2 ] \operatorname{Var}(p_i)
\label{eq:final_var_mix}
\end{align}
where $\zeta_i$ is a vector and thus the square of its expectation is the norm $\bignorm{\mathbb{E}[\zeta_i]}^2$.
The third equality is obtained by applying the identity of the variance
$\operatorname{Var}( \zeta_i ) = \mathbb{E}[\norm{\zeta_i}^2]- \bignorm{\mathbb{E}[\zeta_i]}^2$ to replace $\bignorm{\mathbb{E}[\zeta_i]}^2$.

In contrast, in the standalone training of the prediction component $i$ with the squared error, e.g., training the factor prediction $\hat{r} = g_{\theta_f}(\mathbf{x}_f)$, the stochastic gradient $\delta_i$ and its variance over the training data are expressed as:
\begin{align}
& \delta_i = -2 \zeta_i = -2 (r - \hat{r}) \nabla_{\theta_i} g_{\theta_i}( \mathbf{x}_i )
\\
& \operatorname{Var}( \delta_i ) = 4  \operatorname{Var}(\zeta_i) \label{eq:final_var_standalone}
\end{align}

\end{proof}

\begin{figure}[!htbp]
  \centering
  \begin{subfigure}[b]{0.9\textwidth}
    \includegraphics[width=\textwidth]{figures/na_obs_train.pdf}
    \caption{North American Universe.}
    \label{fig:1}
  \end{subfigure}
  %
  \begin{subfigure}[b]{0.9\textwidth}
    \includegraphics[width=\textwidth]{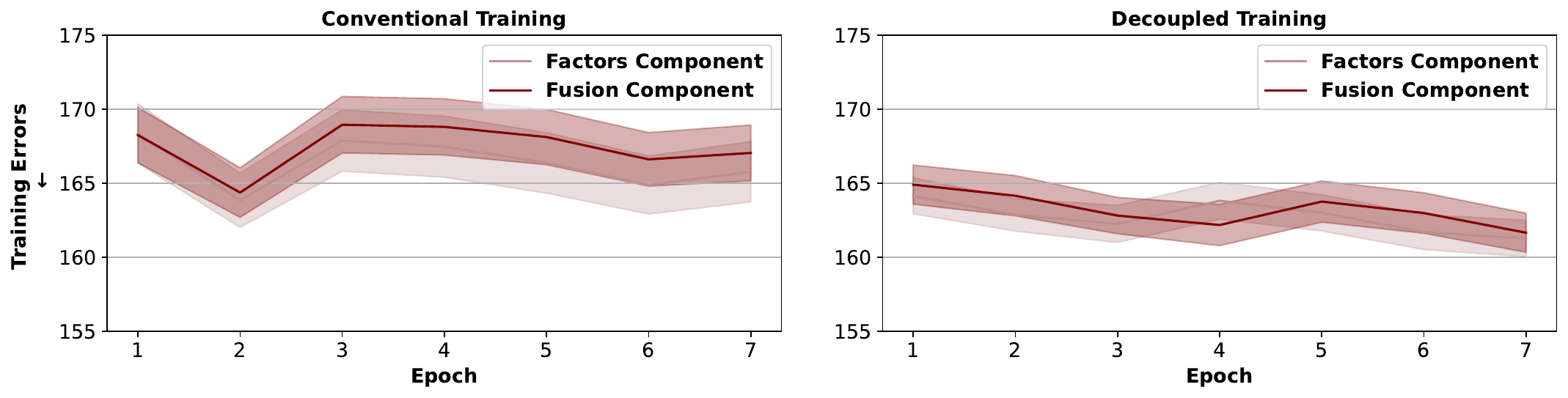}
    \caption{Emerging Markets Universe.}
    \label{fig:2}
  \end{subfigure}
  %
  \begin{subfigure}[b]{0.9\textwidth}
    \includegraphics[width=\textwidth]{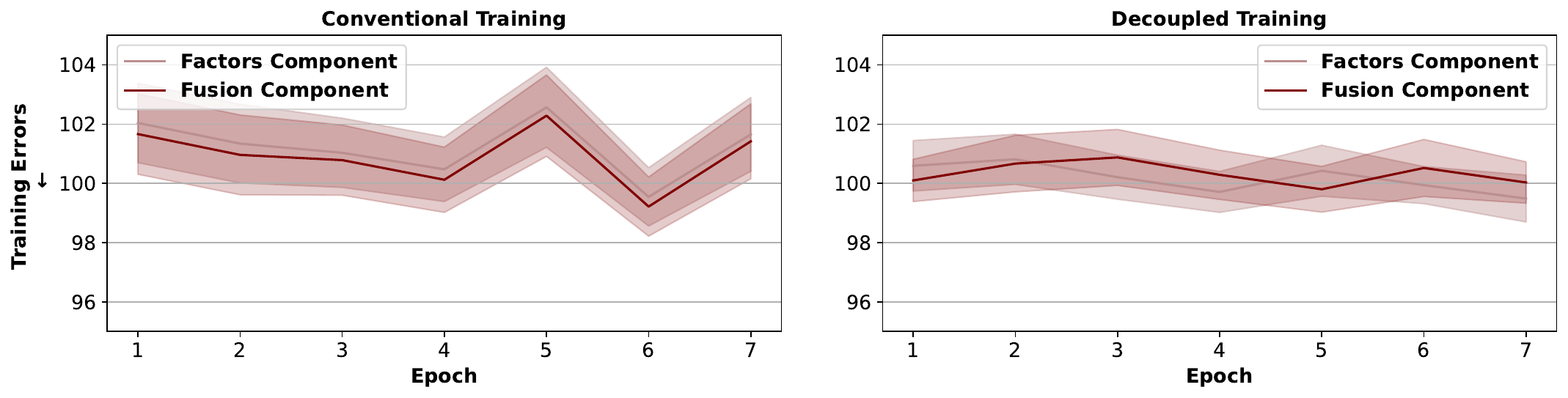}
    \caption{European Universe.}
    \label{fig:2}
  \end{subfigure}
  
  \caption{
  Training error curves of the prediction components in the mixture model under different training methods.
  }
  \label{fig:obs_train}
\end{figure}

\textbf{Discussion.}
For the ease of comparison, in Eq.~\ref{eq:final_var_mix} and Eq.~\ref{eq:final_var_standalone}, we assume a general and approximately comparable $\zeta_i$ across the mixture and standalone training cases.
This assumption can hold when the individual prediction components do not produce drastically different predictions or when $p(I | \cdot )$ might be close to uniform, and individual prediction components are not too specialized during early training.


However, this assumption may not hold exactly, because the residual $r-\hat{r}$ in the standalone training is tied to the corresponding prediction component. 
In the mixture model, the residual depends on the predictions of all components and the probabilities.

Despite this, the proposition remains meaningful because the key result of Proposition 1 is not about the absolute value of the variance but its structural entanglement: the variances of $\zeta_i$ and $p_i$ are multiplicatively coupled with expectation terms in Eq.~\ref{eq:final_var_mix}.
This entanglement introduces additional variance into each prediction component's stochastic update, which leads to instability or convergence issues in training, as shown in Fig.~\ref{fig:obs_train}.

Specifically, in Eq.~\ref{eq:final_var_mix}, the expected squared norm $\mathbb{E}[ \norm{\zeta_i}^2 ]$ can be large due to the number of parameters in prediction components implemented via neural networks.
As a result, even moderate fluctuations in the mixture probability $\operatorname{Var}(p_i)$ can induce amplified variance in the parameter updates, increasing optimization noise and potentially slowing or destabilizing convergence.
Additionally, since the residual $r-\hat{r}$ of the mixture model is influenced by all prediction components, an inaccurate component can inflate the residual, thereby increasing the variance $\operatorname{Var}( \zeta_i )$ in Eq.~\ref{eq:final_var_mix}.


\section{Proof of Proposition 2 and Discussion}

Before presenting the proof, we first define two operators.  
$\operatorname{min}\{a,b\}$ returns the smaller of its two arguments.  
$\operatorname{clip}_{[0,1]}(a)$ returns $1$ if $a>1$, returns $0$ if $a<0$, and returns $a$ otherwise.

\textbf{Proposition 2 (Optimal Mixture Weights and the Relation to the Formulation in Eq.~\eqref{eq:target_distribution}).}
\textit{
For a data instance $(\mathbf{x}_f, \mathbf{x}_n, r)\sim\mathcal{D}$, let $\hat{r}_f$ and $\hat{r}_u$ denote the factors-based and fusion-based predictions under parameters $\hat{\theta}_f$ and $\hat{\theta}_u$.
Define
$
\hat{p}:=\frac{r - \hat{r}_u}{\hat{r}_f - \hat{r}_u}
$
and
$
d:=\hat{r}_f - \hat{r}_u
$.
Then, the optimal mixture weight minimizing the squared prediction error is
$
p_f^* = \operatorname{clip}_{[0,1]}(\hat{p})
$
and
$
p_u^* = 1 - p_f^*
$
, and the mixture prediction
$\hat{r}^* = p_f^* \hat{r}_f + p_u^* \hat{r}_u$
satisfies
$
\mathbb{E}\!\left[(r-\hat{r}^*)^2\right]
\le
\operatorname{min}\!\left\{
\mathbb{E}\!\left[(r-\hat{r}_f)^2\right]
,
\mathbb{E}\!\left[(r-\hat{r}_u)^2\right]
\right\}
$.
Meanwhile, the target distribution defined in Eq.~\eqref{eq:target_distribution} admits an equivalent expression with $\hat{p}$:
$
p_{\hat{\theta}_f,\hat{\theta}_u}(I \mid \mathbf{x}_f, \mathbf{x}_n, r)
=
\left[
\sigma\!\Big( \frac{d^2(2\hat{p}-1)}{\tau} \Big),
\;
1 - \sigma\!\Big( \frac{d^2(2\hat{p}-1)}{\tau} \Big)
\right]
$
, where $\sigma(\cdot)$ is the sigmoid function.
}

\begin{proof}
In the following, we prove the proposition in 3 steps.

\underline{Step 1:}
For a data instance $(\mathbf{x}_f, \mathbf{x}_n, r)\sim\mathcal{D}$, under available parameters $\hat{\theta}_f$ and $\hat{\theta}_u$, let $\hat{r}_f$ and $\hat{r}_u$ denote the factors-based and fusion-based predictions.
Consider the mixture prediction as a function of the mixture weight $p$ as:
\begin{equation}
\hat{r}(p)
:= p\,\hat{r}_f + (1-p)\,\hat{r}_u
= \hat{r}_u + p\,d ,
\qquad d := \hat{r}_f - \hat{r}_u  
\end{equation}

The squared loss is
\begin{equation}
\ell(p) := (r - \hat{r}(p))^2
= (r - \hat{r}_u - p d)^2 
\end{equation}

Taking the derivative $\ell'(p)$ and setting $\ell'(p)=0$ give the unconstrained optimal weight:
\begin{equation}
\hat{p} := \frac{r-\hat{r}_u}{d}
= \frac{r-\hat{r}_u}{\hat{r}_f - \hat{r}_u}
\end{equation}

Since $p$ must lie in $[0,1]$, the valid optimal weight is
\begin{equation}
p_f^* = \operatorname{clip}_{[0,1]}(\hat{p}),
\qquad
p_u^* = 1 - p_f^*
\end{equation}

\underline{Step 2:}
We distinguish two cases for the unconstrained weight $\hat{p}=\dfrac{r - \hat{r}_u}{\hat{r}_f - \hat{r}_u}$.

\textit{(i) Opposite-side case.}
If $(r - \hat{r}_f)(r - \hat{r}_u) < 0$, then $r$ lies between $\hat{r}_f$ and $\hat{r}_u$, hence $\hat{p}\in(0,1)$.
In this case, the quadratic loss $\ell(p) = (r -\hat{r}_u - p(\hat{r}_f - \hat{r}_u))^2$ is minimized at the interior point $p_f^* = p^*$, and by convexity $\ell(p_f^*) < \min\{\ell(0), \ell(1)\}$ (strict unless degenerate). 
Thus, the optimal mixture prediction is an interior convex combination and strictly improves over both endpoints $\hat{r}_f$ and $\hat{r}_u$.

\textit{(ii) Same-side case.}
If $(r - \hat{r}_f)(r - \hat{r}_u) \ge 0$, both predictions lie on the same side of $r$.
Then, the unconstrained optimal weight $\hat{p}$ falls outside $[0,1]$, and the constrained weight is binary, i.e., $p_f^* \in \{0,1\}$.
The clipping operation, therefore, selects the endpoint prediction that is closer to $r$, so $\ell(p_f^*) = \min\{\ell(0), \ell(1)\}$.

Consequently, in all cases
\[
\ell(p_f^*) \le \min\{\ell(0),\ell(1)\},
\]
and taking expectation yields the claimed inequality
\[
\mathbb{E}[(r - \hat{r}^*)^2] \le \min\big\{ \mathbb{E}[(r - \hat{r}_f)^2],\;\mathbb{E}[(r - \hat{r}_u)^2] \big\}.
\]

\underline{Step 3:}
The target distribution defined in Eq.~\eqref{eq:target_distribution} is equivalent to:
\begin{equation}
p_{\hat{\theta}_f,\hat{\theta}_u}(I {=} f \mid \mathbf{x}_f,\mathbf{x}_n,r)
=
\sigma\!\left(\frac{(r-\hat{r}_u)^2-(r-\hat{r}_f)^2}{\tau}\right)
\end{equation}
where $\sigma(\cdot)$ represents a sigmoid function.

Using
\[
(r-\hat{r}_u)^2-(r-\hat{r}_f)^2 = d^2(2\hat{p}-1),
\]
obtained by substituting $\hat{p}=(r-\hat{r}_u)/d$, we obtain an alternative expression of the target distribution in relation to the unconstrained optimal weight $\hat{p}$ as:
\begin{equation}\label{eq:mix_weight_alter}
p_{\hat{\theta}_f,\hat{\theta}_u}(I {=} f \mid \mathbf{x}_f,\mathbf{x}_n,r)
=
\sigma\! \left( \frac{d^2(2\hat{p}-1)}{\tau} \right),
\qquad
p_{\hat{\theta}_f,\hat{\theta}_u}(I {=} u \mid \cdot) = 1 - p_{\hat{\theta}_f, \hat{\theta}_u}(I {=} f \mid \cdot)
\end{equation}
which completes the proof.

\end{proof}

\textbf{Discussion.}
In Eq.~\ref{eq:mix_weight_alter}, the sign of $2\hat{p} - 1$ indicates which component prediction is favored. 
Specifically, $2\hat{p} - 1 > 0$ leads to a higher weight for the factor-based prediction, and otherwise, $2\hat{p} - 1 < 0$, the fusion-based prediction has a higher weight.
Thus, the sign of $2\hat{p} - 1$ preserves the same ordering of mixture weights as the optimal mixture weight.
The magnitude of $d^{2}$ governs the extent of the overweight or underweight: when $d^{2}$ is large, it pushes the sigmoid output closer to $0$ or $1$, mirroring the fact that the optimal mixture weight becomes binary when both predictions lie on the same side of the true value.

Overall, Eq.~\ref{eq:target_distribution} can be viewed as a monotone transformation of the signal $\hat{p}$ with the additional adaptive scaling $d^{2}$.
At the same time, unlike the optimal weight, which can be binary, the target distribution defined in Eq.~\ref{eq:target_distribution} remains relatively smooth, which helps stabilize optimization and mitigates overfitting.


\section{Theoretical Insights of Decoupled Training}\label{appendix:theory}

In this part, we discuss the implications of the decoupled training formulation and its connection to variational inference~\cite{blei2017variational, zhang2018advances}, emphasizing its theoretical grounding and practical relevance.

In the decoupled training, we formulate the distribution matching as the KL divergence between the trainable distribution $p_{\phi}(I \,| \mathbf{x}_{f}, \mathbf{x}_n )$ and the target distribution $p_{\hat{\theta}_f, \hat{\theta}_u}(I \,| \mathbf{x}_f, \mathbf{x}_n, r )$ realized with the parameter estimates $ \hat{\theta}_f$ and $\hat{\theta}_u$, as shown in Eq.~\ref{eq:loss_prob}:
\begin{align*}
L\big( \phi \,; \{ \mathbf{x}_f, \mathbf{x}_n, r \}, \hat{\theta}_f, \hat{\theta}_u \big) 
:=
\kl\Big[
p_{\phi}(I \,| \mathbf{x}_{f}, \mathbf{x}_n ) 
\,\big\Vert\,
p_{ \hat{\theta}_f, \hat{\theta}_u}(I \,| \mathbf{x}_f, \mathbf{x}_n, r )
\Big] 
\tag{\ref{eq:loss_prob}}
\end{align*}

KL divergence is not symmetric, and minimizing the KL divergence like Eq.~\ref{eq:loss_prob} encourages the trainable distribution $p_{\phi}(I \,| \mathbf{x}_{f}, \mathbf{x}_n )$ to concentrate on modes of the target distribution $p_{\hat{\theta}_f, \hat{\theta}_u}(I \,| \mathbf{x}_f, \mathbf{x}_n, r )$, as it penalizes the assignment of probability mass to regions where the target distribution has low density.
This helps make the model robust to noisy or irrelevant inputs.

Next, we interpret the decoupled learning loss function in Eq.~\ref{eq:loss_decoupled} through the lens of variational inference, where a KL divergence is used to align an approximate posterior with a true posterior distribution.

We reformulate the return prediction model from the perspective of probabilistic latent variable models as: 
\begin{align}\label{eq:prob_model}
p( r | \mathbf{x}_f, \mathbf{x}_n ) 
= 
\sum_{i \in \{f, u\} } {p}( I=i ) \, p( r | I = i, \mathbf{x}_i ) 
\end{align}

Eq.~\ref{eq:prob_model} takes $I$ as a latent random variable.
$p( r | I = i, \mathbf{x}_i )$ corresponds to the prediction component in our mixture model.
${p}( I=i )$ is a prior over $I$.

Applying variational inference to a latent variable model such as Eq.~\ref{eq:prob_model} seeks to maximize the evidence lower bound (ELBO) on the log likelihood~\cite{blei2017variational, zhang2018advances}, as shown below:
\begin{align}
& \log p( r | \mathbf{x}_f, \mathbf{x}_n ) \\
& \geq \log p( r | \mathbf{x}_f, \mathbf{x}_n ) 
-
\kl\big[ q_{}( I \,| \mathbf{x}_{f}, \mathbf{x}_n ) \,\big\Vert\, p_{}(I \,| \mathbf{x}_f, \mathbf{x}_n, r ) \big] 
\label{eq:vi_loss}
\\
& = 
\underbrace{
\sum_{i \in \{f, u\}}  q_{}( I=i \,| \mathbf{x}_{f}, \mathbf{x}_n ) \log p( r | I = i, \mathbf{x}_i)
-
\kl\big[ q_{}( I \,| \mathbf{x}_{f}, \mathbf{x}_n ) \,\big\Vert\, p_{}(I) \big]
}_{
\text{ELBO}
}
\end{align}
In Eq.~\ref{eq:vi_loss}, the posterior distribution $p_{}(I \,| \mathbf{x}_f, \mathbf{x}_n, r )$ reflects the relative performance of each prediction component given the true return $r$ and $\mathbf{x}_f$ and $\mathbf{x}_n$.
$q_{}( I \,| \mathbf{x}_{f}, \mathbf{x}_n )$ is the trainable distribution for approximating the true posterior distribution $p_{}(I \,| \mathbf{x}_f, \mathbf{x}_n, r )$ of the latent variable $I$.
At test time, the true posterior $p_{}(I \,| \mathbf{x}_f, \mathbf{x}_n, r )$ is unknown, and $q_{}( I \,| \mathbf{x}_{f}, \mathbf{x}_n )$ servers to infer the posterior.

In parallel, we present a lower bound of Eq.~\ref{eq:vi_loss}, which leads to our decoupled training loss under certain assumptions.
\begin{align}
& \log p( r | \mathbf{x}_f, \mathbf{x}_n ) 
-
\kl\big[ q_{}( I \,| \mathbf{x}_{f}, \mathbf{x}_n ) \,\big\Vert\, p_{}(I \,| \mathbf{x}_f, \mathbf{x}_n, r ) \big] 
\tag{\ref{eq:vi_loss}}
\\
& = 
\log \sum_{i \in \{f, u\}} {p}( I=i ) \, p( r | I = i, \mathbf{x}_i)
-
\kl\big[ q_{}( I \,| \mathbf{x}_{f}, \mathbf{x}_n ) \,\big\Vert\, p_{}(I \,| \mathbf{x}_f, \mathbf{x}_n, r ) \big]
\\
& \geq 
\sum_{i \in \{f, u\}} {p}( I=i ) \log p( r | I = i, \mathbf{x}_i)
-
\kl\big[ q_{}( I \,| \mathbf{x}_{f}, \mathbf{x}_n ) \,\big\Vert\, p_{}(I \,| \mathbf{x}_f, \mathbf{x}_n, r ) \big] 
\label{eq:vi_diff}
\end{align}
The inequality in Eq.~\ref{eq:vi_diff} holds due to the concavity of the logarithm function.

Equivalently, in practice, we typically minimize the negative log-likelihood for training the model, which corresponds to minimizing the negative of Eq.~\ref{eq:vi_diff} as follows:
\begin{align}
\underbrace{
-\sum_{i \in \{f, u\}} {p}( I=i ) \log p( r | I = i, \mathbf{x}_i)
}_{ \mathcal{L}( \cdot ) }
+
\underbrace{
\kl\big[ q_{}( I \,| \mathbf{x}_{f}, \mathbf{x}_n ) \,\big\Vert\, p_{}(I \,| \mathbf{x}_f, \mathbf{x}_n, r ) \big] 
}_{ \text{KL}(\cdot) }
\label{eq:vi_mini}
\end{align}

In the following, we show that, under certain assumptions, the two terms in Eq.~\ref{eq:vi_mini} correspond to those in the decoupled training objective Eq.~\ref{eq:loss_decoupled}.

Assuming that $p( r | I = i, \mathbf{x}_i)$ follows a Gaussian distribution with constant variance, minimizing the negative log likelihood $-\log p( r | I = i, \mathbf{x}_i)$ reduces to minimizing the squared error between the true value and predictive mean of the distribution, i.e., $\big[ r - g(x) \big]^2$.
Meanwhile, if the prior ${p}( I=i )$ is not trainable, then
\begin{align}
\underset{\theta_i}{\argmin} \, \mathcal{L}( \cdot ) 
= 
\underset{\theta_i}{\argmin} 
-\log p_{\theta_i}( r | \mathbf{x}_i, I = i)
= 
\underset{\theta_i}{\argmin}
\big[ r - g_{\theta_i}(x) \big]^2
\end{align}

Consequently, the term $\mathcal{L}( \cdot )$ in Eq.~\ref{eq:vi_mini} amounts to the independent training term in Eq.~\ref{eq:loss_decoupled}.

The trainable approximate posterior $q_{}( I \,| \mathbf{x}_{f}, \mathbf{x}_n )$ is equivalent to the trainable distribution $p_{\phi}(I \,| \mathbf{x}_{f}, \mathbf{x}_n )$ used in the distribution matching term Eq.~\ref{eq:loss_prob} of the decoupled training.
The target distribute $p_{ \hat{\theta}_f, \hat{\theta}_u}(I \,| \mathbf{x}_f, \mathbf{x}_n, r )$ in Eq.~\ref{eq:loss_prob} instantiates the posterior distribution $p_{}(I \,| \mathbf{x}_f, \mathbf{x}_n, r )$ in Eq.~\ref{eq:vi_mini}, as the posterior distribution reduces to Eq.~\ref{eq:target_distribution} under a uniform prior of ${p}( I=i )$. 
Overall, the distribution matching in Eq.~\ref{eq:loss_prob} corresponds to the $\text{KL}(\cdot)$ term in the variational inference objective.


\section{Experiment Details and Full Results}


\subsection{Datasets}

Table~\ref{tab:factors_categories} presents the main categories of these factors.
Our factors are grounded in financial theories and capture a range of stock characteristics.
The total number of factors is $\sim200$.
\begin{table}[htbp]
\caption{Quantitative Factor Categories}
\centering
\begin{tabular}{|p{11cm}|}
\hline
Quality, Size, Volume, Growth, Momentum, Risk, Revision, Valuation, etc. \\
\hline
\end{tabular}
\label{tab:factors_categories}
\end{table}

Table~\ref{tab:news_categories} lists the main categories of news included in our dataset.
These category labels are provided directly by the data vendor alongside news data.
Since our focus is on return prediction for stock selection, we primarily use news data that covers company-specific events such as earnings reports, ratings, outlooks, corporate actions, etc.
Each piece of news has an attribute including the company identifier(s) the news is linked to.
\begin{table}[h]
\caption{News Categories}
\centering
\begin{tabular}{|p{13cm}|}
\hline
Earnings, Guidance, Upgrades, Downgrades, Mergers and Acquisitions, Corporate Actions, Restructuring, Jobs, Ownerships, Short Interests, Buybacks, Equity Offerings, Management Changes, etc. \\
\hline
\end{tabular}
\label{tab:news_categories}
\end{table}

For associating news with stocks, we use a one-week look-back window for the EU and NA universes and a one-month window for the EM universe.
The longer window for EM is due to its relatively lower news coverage.
Then, for each data instance, we construct the newsflow using headlines from our news datasets, as the data, sourced from a professional financial vendor, provides structured and concise headlines~\cite{chen2021stock, lopez2023can}. 
The full article content is noisier and introduces substantially higher training overhead, likely requiring a relevance-filtering step to retain only relevant information.
Therefore, we leave the exploration of full content to future work.

The training and validation data span from 2003 to 2022, and the testing data cover the two-year period from 2023 to 2024.
The volume of news data fluctuates annually, with a notable increase in recent years.
Since the LLM used in this paper, DeBERTa, was developed around 2022, we selected 2023-2024 as the testing period to mitigate potential memorization bias from the LLM~\cite{rahimikia2024r, levy2024caution,lopez2025memorization}.

Table~\ref{tab:data_size} presents the stats of training, validation, and testing data.
The Range of News Items column indicates the number of news items in the newsflow associated with each data instance, and the Range of Tokens shows the number of tokens resulting from tokenizing the newsflow of each data instance.
Our investment universes include all-cap stocks, and the coverage of news data is modest relative to these universes.
Consequently, there exist instances with only one news term, resulting in a low token count, sometimes fewer than 10, as shown in Table~\ref{tab:data_size}.
\begin{table*}[htbp]
\centering
\caption{
Stats of Training, Validation, and Testing Data. 
}
\resizebox{1.\textwidth}{!}{
\begin{tabular}{|c|c|c|c|c|c|c|}
\hline
Universe & \# of Stocks  & \# of Training Instances & \# of Validating Instances & \# of Testing Instances & Range of News Items & Range of Tokens\\
\hline
North America     & 830      & 634214 & 10167 & 270345  &  [1, 98]  & [8, 1866] \\
Emerging Markets  & 1090     & 213830 & 10183 & 263380  &  [1, 96]  & [6, 1725] \\
Europe            & 370      & 201863 & 10094 & 113303  &  [1, 51]  & [5, 914] \\
\hline
\end{tabular}
}
\label{tab:data_size}
\end{table*}


\subsection{Baselines}
\underline{Universe} refers to a portfolio that equally weights all stocks in the investment universe. 
Its performance is reported in the Long-Only Portfolio section.

\underline{Factors Alone} represents a multiple dense neural network solely on quantitative factors~\cite{chen2024deep}.
Note that for a fair comparison, in our mixture model, the factors-based prediction component adopts the same model structure as this baseline.

\underline{News Alone} utilizes only news by applying a prediction layer on the news representations generated by an LLM~\cite{guo2024fine}.
Specifically, it uses the encoder-only LLM, i.e., DeBERTa, consistent with the model used in our fusion learning and mixture modeling approaches.
We report the performance without and with enabling the fine-tuning of DeBERTa during training.

\underline{FININ} is close to the representation combination approach in our fusion learning framework, but differs in that it uses market data (e.g., factors in this paper) to weight news representations before jointly passing both through a prediction layer~\cite{wang2024modeling}.


\subsection{Implementations}

For a fair comparison, we employ the encoder-only LLM, DeBERTa~\cite{he2021debertav3}, across our fusion learning methods, mixture models, and all LLM-based baselines.
DeBERTa improves upon encoder-only language models using disentangled content and position embeddings and has demonstrated competitive performance in various financial tasks~\cite{guo2024fine, muhammad2025assessing}.
In contrast, larger decoder-only LLMs such as Mistral and LLaMA may risk memorizing market information during pre-training on large and recent datasets, potentially introducing bias into downstream evaluations~\cite{lopez2025memorization, engelberg2025entity}.

All the methods below are implemented with the PyTorch~\cite{paszke2019pytorch} and HuggingFace Transformers libraries~\cite{wolf2019huggingface}.

\underline{Fusing Learning.}
The representation combination method is implemented by first concatenating factor features with a bottleneck representation of newsflow, which is then passed through a dense layer for fusion, followed by a linear output layer~\cite{zhang2025camef, chen2020uniter}.
We empirically find that reducing the dimensionality of the newsflow representation, using a bottleneck dense layer, generally improves performance.
This is implemented by a single dense layer that compresses the LLM-generated news representation to half of its original dimension.

For the representation summation and attentive representation methods, we use two single-layer dense networks to respectively project the factor and news representations into a unified representation space~\cite{kiela2018efficient}.
To ensure fair comparison, the dimension of this unified space is matched to the output dimension of the fusion dense layer used in the representation combination method.

In the representation summation approach, the two projected representations are summed and passed directly to a linear output layer.
In the attentive representation approach, the factor and news representations are used to compute attention logits via a dense layer~\cite{nagrani2021attention, koval2024financial}.
These logits define the weighting over the two projected representations, and the resulting weighted sum is passed to the output layer.

Note that the attentive representation method is conceptually closely related to the self-attention mechanism in Transformers~\cite{vaswani2017attention}.
The weight vector learned in the dense layer can be interpreted as a query vector in self-attention, and thus the attention weights $a_f$ and $a_n$ in the attentive representation method represent the relevance of factors and news representations to this query.

\underline{Mixture Modeling.}
For mixture modeling, the fusion component is implemented as described in the representation combination method.
The factor component is realized using two dense layers with skip connections~\cite{gu2020empirical}.
The probability logits function is implemented by a dense layer on factors and news representations.

\underline{Baselines.}
The Factors-alone baseline follows the same architecture as the factor component in the mixture model.
The News-alone baseline uses a linear output layer on top of aggregated token-level representations, following the approach in~\cite{guo2024fine}.

Following~\cite{wang2024modeling}, the FININ baseline computes the weight through the scaled dot product of two representation vectors, which respectively correspond to factors and news.
These two representations are obtained by respectively passing the factor and news embeddings through two separate single-layer dense networks.
Then, the weighted representation of news is added to the factors' representation and fed to the output layer.


\subsection{Training and Evaluation Setup}

For training, we use the one-month forward return as the target variable, as the subsequent backtest focuses on monthly rebalanced portfolios.
We conduct the training using a batch size $32$ and a learning rate $1e-4$ with a linear decay scheduler.
For regularization, we apply a dropout rate of $0.3$ to the input of prediction layers and set the weight decay to $1e-4$.
After training, the model is evaluated on the testing period without retraining in a rolling manner.


For fine-tuning, we applied Low-Rank Adaptation (LoRA) with rank $4$ to all linear layers~\cite{hu2021lora}.
Other techniques, including gradient checkpointing, mixed precision training, and DeepSpeed, are used to reduce GPU memory~\cite{rasley2020deepspeed}.
We employ a maximum context length of $4$k in experiments. 
All models are trained for $10$ epochs on $2\times\text{A}100$ GPUs.

During backtesting, the long-only portfolio is constructed by selecting the stocks whose return predictions fall in the top (9th) decile of the prediction rankings.
The long-short portfolio includes stocks from both the top (9th) and bottom (0th) deciles.
All stocks within each portfolio are equally weighted, and both portfolios are rebalanced on a monthly basis.


\subsection{Metrics}

For portfolio performance, we report annualized returns and Sharpe ratios over the testing period, along with charts of cumulative returns.
Additionally, we present the bar charts of decile returns to provide insights into the sources of portfolio performance.

\underline{Decile Return}, or the return per predicted decile, is derived in the following way~\cite{guo2024fine}.
At each rebalancing date, stocks are sorted by predicted returns and grouped into 10 deciles, labeled $d = 0, \dots, 9$.
For each decile $d$, we calculate the average return of the stocks within that decile.
These decile-level returns are then aggregated over the testing period to obtain the final decile return profile.

We also report two prediction performance metrics.
\underline{Mean Absolute Percentage Error (MAPE)} measures the average of the absolute percentage differences between predicted and actual values.
We chose MAPE over RMSE or MSE because it expresses the error relative to the actual value, making it less sensitive to outliers and useful for comparing errors across value scales.

\underline{Information Coefficient (IC)} quantifies the rank correlation between predicted and actual values~\cite{duan2022factorvae}. 
It is particularly relevant for return prediction tasks, as stock selection often depends on the ranking of predicted returns. 
A high IC indicates strong alignment between the predicted and actual rankings, which typically leads to better portfolio performance.


\clearpage
\subsection{Results of the North American Universe}

\begin{table*}[!htbp]
  \centering
  \caption{
  Portfolio and prediction performance of the North American Universe.
  The best and second-best results are highlighted with dark gray \rankone{\,\,} and light gray \ranktwo{\,\,} boxes, respectively.
  }
  \resizebox{1.\textwidth}{!}{
  \begin{tabular}{|c|c|c|c|c|c|c|}
    \hline
    \multirow{2}{*}{} & \multicolumn{2}{|c|}{Long-only Portfolios} & \multicolumn{2}{|c|}{Long-short Portfolios} & \multicolumn{2}{|c|}{Prediction Metrics}\\
    \cline{2-7}
    & Ann. Return \% ($\uparrow$) & Sharpe Ratio ($\uparrow$) & Ann. Return \% ($\uparrow$) & Sharpe Ratio ($\uparrow$) & MAPE ($\downarrow$) & IC ($\uparrow$) \\
    \hline
Universe       & 12.37 & 0.84 & $-$  & $-$    & $-$ & $-$  \\
Factors Alone & 22.31 & 0.81 & 22.34 & 1.26 & 1.352 & 0.018 \\
News Alone & 20.96 & \rankone{1.03} & 1.08 & 0.18 & \rankone{1.092} & -0.0 \\
FININ & 23.12 & 0.83 & 18.16 & 1.16 & 1.467 & 0.019 \\
\hline
Fusion Combination & \rankone{32.43} & \ranktwo{1.0} & 28.41 & \ranktwo{1.64} & 1.402 & \rankone{0.031} \\
Fusion Summation & 20.42 & 0.76 & 16.13 & 1.03 & 1.465 & 0.017 \\
Fusion Attention & 21.73 & 0.73 & 19.59 & 0.87 & \ranktwo{1.302} & -0.005 \\
\hline
Mixture Conventional & 23.27 & 0.75 & \ranktwo{29.32} & 1.42 & 1.539 & 0.016 \\
Mixture Decoupled & \ranktwo{28.21} & 0.92 & \rankone{33.77} & \rankone{1.78} & 1.319 & \ranktwo{0.027} \\
  \hline
  \end{tabular}
  }
  \label{tab:appendix_na_portfolio}
\end{table*}

\begin{figure}[h]
  \begin{center}
    \includegraphics[width=1.0\textwidth]{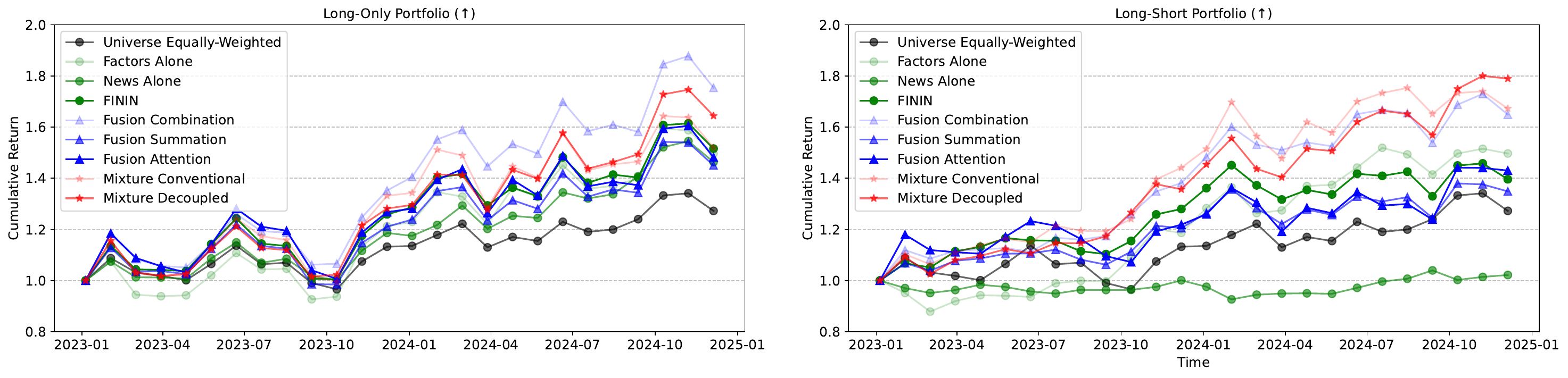}
  \end{center}
  \caption{
  Portfolio Performance Charts of the North American Universe.
  }
  \label{fig:appendix_na_charts}
\end{figure}

\begin{figure}[!htbp]
  \begin{center}
    \includegraphics[width=0.7\textwidth]{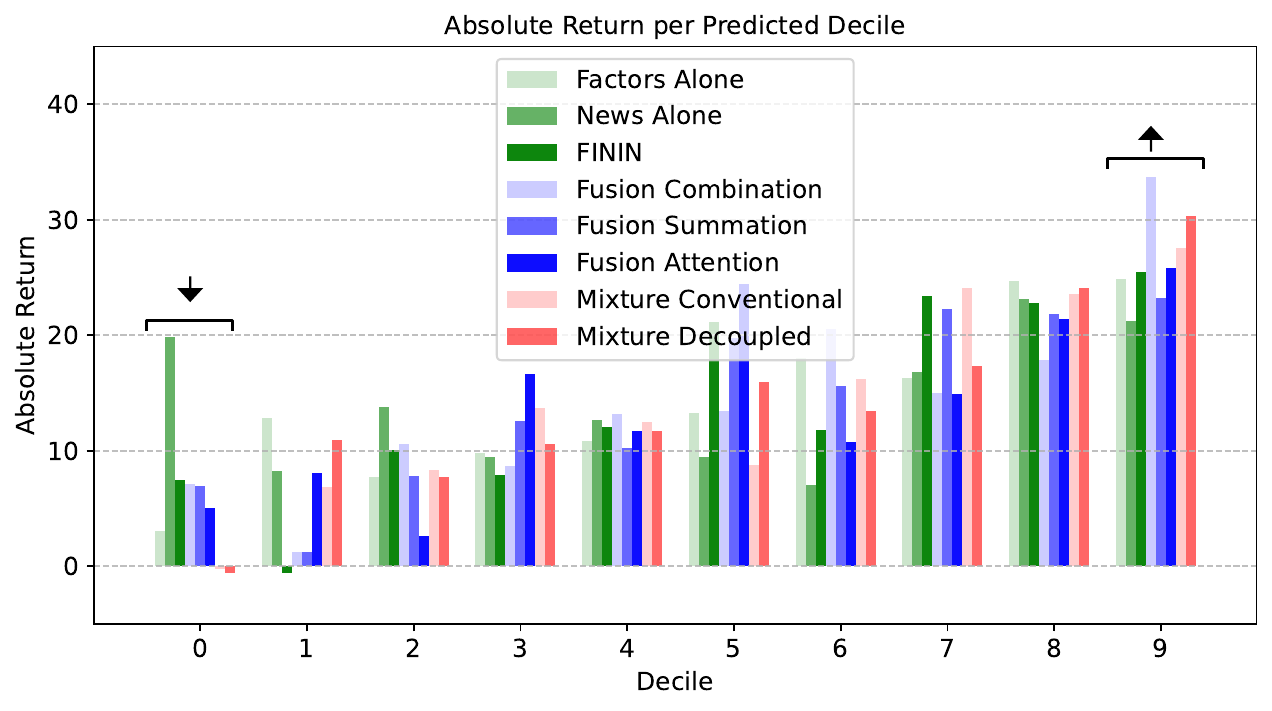}
  \end{center}
  \caption{
  Decile Returns of the North American Universe.
  The arrows on the 0th and 9th deciles indicate the desired direction of values.
  A lower return is preferred for the 0th decile, as it represents the short leg of a long-short portfolio.
  }
  \label{fig:appendix_na_deciles}
\end{figure}

\begin{table*}[t]
  \centering
  \caption{
  Portfolio and prediction performance without and with enabling LLM fine-tuning during training for the North American Universe.
  The best and second-best results within each group (without fine-tuning and with fine-tuning) are highlighted with dark gray \rankone{\,\,} and light gray \ranktwo{\,\,} boxes. 
  }
  \resizebox{1.\textwidth}{!}{
  \begin{tabular}{|c|c|c|c|c|c|c|c|}
    \hline
    \multirow{2}{*}{} &  & \multicolumn{2}{|c|}{Long-only Portfolios} & \multicolumn{2}{|c|}{Long-short Portfolios} & \multicolumn{2}{|c|}{Prediction Metrics}\\
    \cline{3-8}
    & & Ann. Return \% ($\uparrow$) & Sharpe Ratio ($\uparrow$) & Ann. Return \% ($\uparrow$) & Sharpe Ratio ($\uparrow$) & MAPE ($\downarrow$) & IC ($\uparrow$) \\
    \hline
\multirow{7}{*}{\rotatebox{90}{w/o Fine-tuning}}
&News Alone & 20.96 & \rankone{1.03} & 1.08 & 0.18 & \rankone{1.092} & -0.0 \\
&FININ & 23.12 & 0.83 & 18.16 & 1.16 & 1.467 & 0.019 \\
&Fusion Combination & \rankone{32.43} & \ranktwo{1.0} & 28.41 & \ranktwo{1.64} & 1.402 & \rankone{0.031} \\
&Fusion Summation & 20.42 & 0.76 & 16.13 & 1.03 & 1.465 & 0.017 \\
&Fusion Attention & 21.73 & 0.73 & 19.59 & 0.87 & \ranktwo{1.302} & -0.005 \\
&Mixture Conventional & 23.27 & 0.75 & \ranktwo{29.32} & 1.42 & 1.539 & 0.016 \\
&Mixture Decoupled & \ranktwo{28.21} & 0.92 & \rankone{33.77} & \rankone{1.78} & 1.319 & \ranktwo{0.027} \\
\hline
\multirow{7}{*}{\rotatebox{90}{w/ Fine-tuning}}
&News Alone & \ranktwo{27.98} & \rankone{1.33} & 14.23 & 1.52 & \rankone{1.118} & 0.018 \\
&FININ & 22.09 & 0.81 & 19.65 & 1.24 & 1.468 & 0.02 \\
&Fusion Combination & \rankone{28.61} & \ranktwo{0.95} & 26.82 & \ranktwo{1.63} & 1.379 & \rankone{0.032} \\
&Fusion Summation & 20.69 & 0.77 & 18.31 & 1.12 & 1.477 & 0.017 \\
&Fusion Attention & 17.12 & 0.63 & 13.55 & 0.7 & \ranktwo{1.298} & -0.004 \\
&Mixture Conventional & 26.57 & 0.85 & \rankone{32.23} & 1.43 & 1.326 & 0.017 \\
&Mixture Decoupled & 27.15 & 0.91 & \ranktwo{30.66} & \rankone{1.79} & 1.336 & \ranktwo{0.028} \\
    \hline
  \end{tabular}
  }
  \label{tab:appendix_na_finetune}
\end{table*}

\begin{figure}[hb]
  \begin{center}
    \includegraphics[width=1.\textwidth]{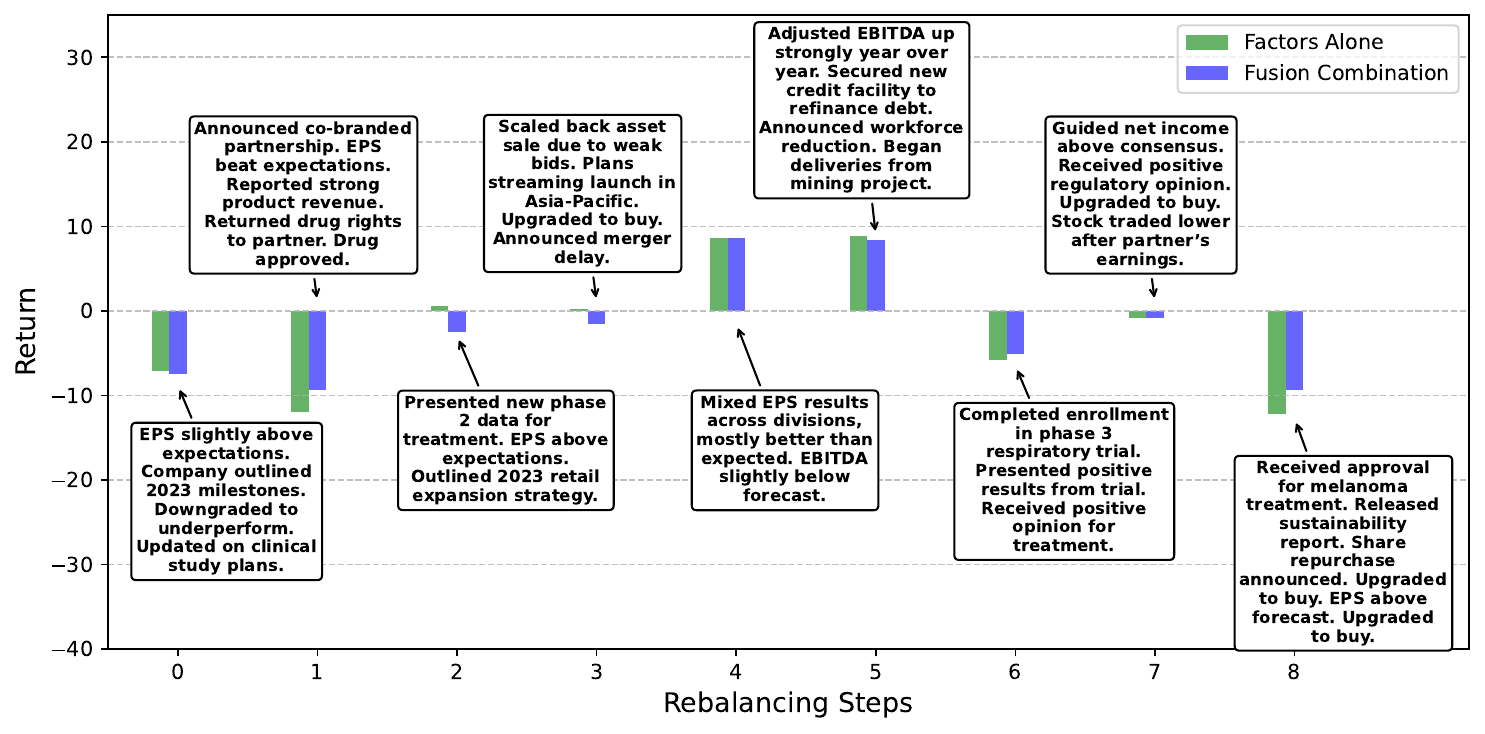}
  \end{center}
  \caption{
  Qualitative Illustration of News Relevance at Rebalancing Steps
  in the North American Universe.
  At a given rebalancing step, the newsflow associated with stocks in the long-only portfolio, built based on return predictions, is collected and summarized for visualization in text boxes. 
  The accompanying bar chart presents the forward returns of portfolios constructed using predictions from the \textit{Factors Alone} and \textit{Fusion Combination} methods at each rebalancing step.
  News contributes positively when it provides relevant and complementary information beyond what is captured by quantitative factors. 
  Conversely, it can be detrimental when the information from the news is irrelevant or already reflected in the factors.
  \\
  \textit{Observations:}
  For instance, at steps 0, 2, and 3, news mostly concerns earnings, sales, and ratings, which tend to carry overlapping information with the growth, quality, and other factors.
  As a result, \textit{Fusion Combination} exhibits weaker performance relative to \textit{Factors Alone}.
  Conversely, at steps 1, 6, and 8, news covers diverse topics such as brand partnerships, trial completion, and so on; consequently, \textit{Fusion Combination} performs competitively. 
  }
  \label{fig:appendix_na_visual_text}
\end{figure}
%
%


\clearpage
\subsection{Results of the Emerging Markets Universe}

\begin{table*}[!htbp]
  \centering
  \caption{
  Portfolio and prediction performance of the Emerging Markets Universe.
  The best and second-best results are highlighted with dark gray \rankone{\,\,} and light gray \ranktwo{\,\,} boxes, respectively.
  }
  \resizebox{1.\textwidth}{!}{
  \begin{tabular}{|c|c|c|c|c|c|c|}
    \hline
    \multirow{2}{*}{} & \multicolumn{2}{|c|}{Long-only Portfolios} & \multicolumn{2}{|c|}{Long-short Portfolios} & \multicolumn{2}{|c|}{Prediction Metrics}\\
    \cline{2-7}
    & Ann. Return \% ($\uparrow$) & Sharpe Ratio ($\uparrow$) & Ann. Return \% ($\uparrow$) & Sharpe Ratio ($\uparrow$) & MAPE ($\downarrow$) & IC ($\uparrow$) \\
    \hline
Universe       & 2.63 & 0.24 & $-$  & $-$    & $-$ & $-$  \\
Factors Alone & \ranktwo{17.14} & \ranktwo{0.81} & \rankone{42.17} & \rankone{2.96} & 1.461 & 0.049 \\
News Alone & -3.94 & -0.2 & -9.67 & -1.57 & \rankone{1.194} & -0.015 \\
FININ & 12.9 & 0.72 & 30.02 & 2.15 & 1.445 & 0.046 \\
\hline
Fusion Combination & 13.36 & 0.75 & 32.35 & 2.21 & 1.452 & \ranktwo{0.06} \\
Fusion Summation & 13.38 & 0.74 & 28.52 & 1.96 & 1.474 & 0.043 \\
Fusion Attention & 11.85 & 0.67 & 14.22 & 1.07 & \ranktwo{1.283} & 0.003 \\
\hline
Mixture Conventional & 7.71 & 0.46 & 22.47 & 1.53 & 1.465 & 0.053 \\
Mixture Decoupled & \rankone{18.5} & \rankone{0.94} & \ranktwo{42.07} & \ranktwo{2.93} & 1.395 & \rankone{0.065} \\
  \hline
  \end{tabular}
  }
  \label{tab:appendix_em_portfolio}
\end{table*}

\begin{figure}[!htbp]
  \begin{center}
    \includegraphics[width=1.0\textwidth]{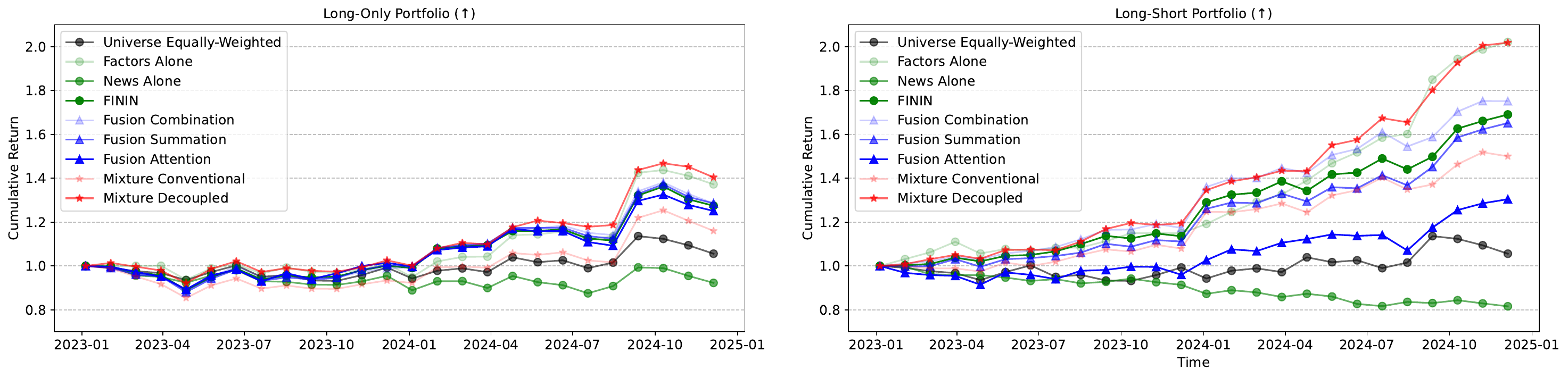}
  \end{center}
  \caption{
  Portfolio Performance Charts of the Emerging Markets Universe.
  }
  \label{fig:appendix_em_charts}
\end{figure}

\begin{figure}[!htbp]
  \begin{center}
    \includegraphics[width=0.7\textwidth]{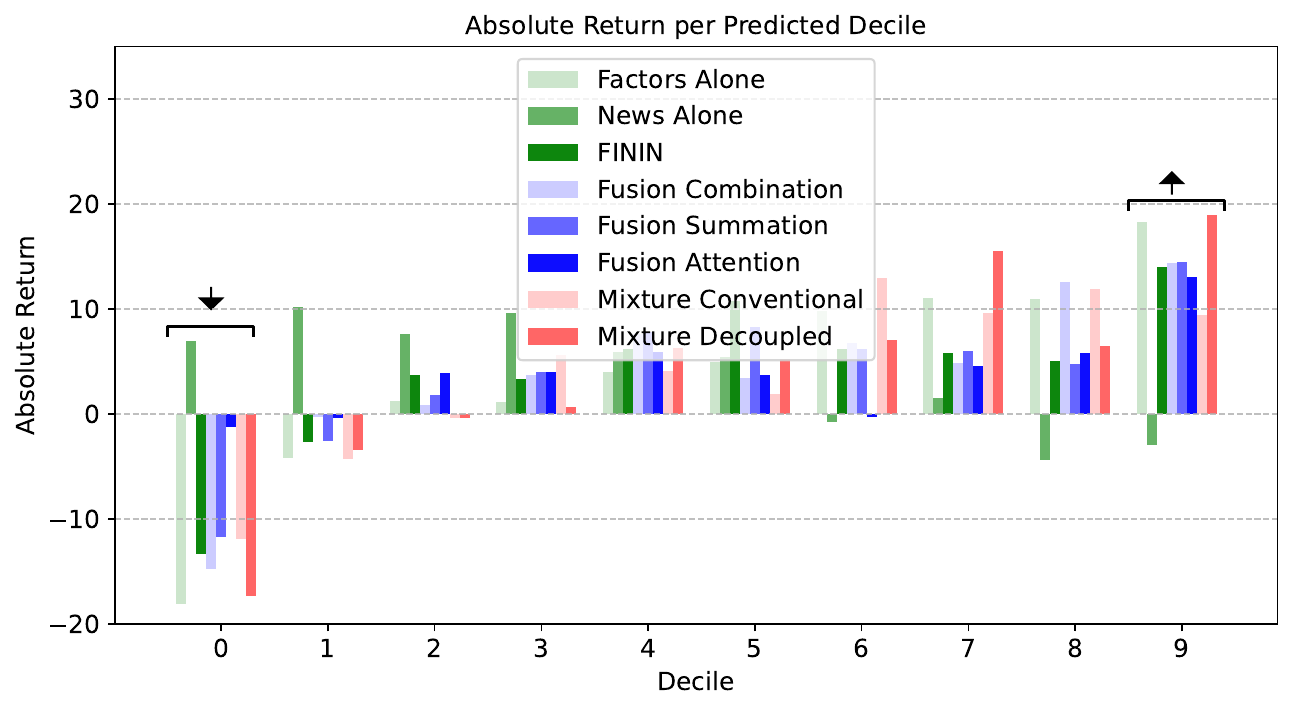}
  \end{center}
  \caption{
  Decile Returns of the Emerging Markets Universe.
  The arrows on the 0th and 9th deciles indicate the desired direction of values.
  A lower return is preferred for the 0th decile, as it represents the short leg of a long-short portfolio.
  }
  \label{fig:appendix_em_deciles}
\end{figure}

\begin{table*}[!htbp]
  \centering
  \caption{
  Portfolio and prediction performance without and with enabling LLM fine-tuning during training for the Emerging Markets Universe.
  The best and second-best results within each group (without fine-tuning and with fine-tuning) are highlighted with dark gray \rankone{\,\,} and light gray \ranktwo{\,\,} boxes. 
  }
  \resizebox{1.\textwidth}{!}{
  \begin{tabular}{|c|c|c|c|c|c|c|c|}
    \hline
    \multirow{2}{*}{} &  & \multicolumn{2}{|c|}{Long-only Portfolios} & \multicolumn{2}{|c|}{Long-short Portfolios} & \multicolumn{2}{|c|}{Prediction Metrics}\\
    \cline{3-8}
    & & Ann. Return \% ($\uparrow$) & Sharpe Ratio ($\uparrow$) & Ann. Return \% ($\uparrow$) & Sharpe Ratio ($\uparrow$) & MAPE ($\downarrow$) & IC ($\uparrow$) \\
    \hline
\multirow{7}{*}{\rotatebox{90}{w/o Fine-tuning}}
&News Alone & -3.94 & -0.2 & -9.67 & -1.57 & \rankone{1.194} & -0.015 \\
&FININ & 12.9 & 0.72 & 30.02 & 2.15 & 1.445 & 0.046 \\
&Fusion Combination & 13.36 & \ranktwo{0.75} & \ranktwo{32.35} & \ranktwo{2.21} & 1.452 & \ranktwo{0.06} \\
&Fusion Summation & \ranktwo{13.38} & 0.74 & 28.52 & 1.96 & 1.474 & 0.043 \\
&Fusion Attention & 11.85 & 0.67 & 14.22 & 1.07 & \ranktwo{1.283} & 0.003 \\
&Mixture Conventional & 7.71 & 0.46 & 22.47 & 1.53 & 1.465 & 0.053 \\
&Mixture Decoupled & \rankone{18.5} & \rankone{0.94} & \rankone{42.07} & \rankone{2.93} & 1.395 & \rankone{0.065} \\
\hline
\multirow{7}{*}{\rotatebox{90}{w/ Fine-tuning}}
&News Alone & 0.38 & 0.1 & 6.43 & 0.87 & \rankone{1.13} & 0.02 \\
&FININ & 13.4 & 0.76 & \ranktwo{31.53} & \ranktwo{2.3} & 1.448 & 0.048 \\
&Fusion Combination & 12.89 & 0.74 & 30.18 & 2.0 & 1.466 & \ranktwo{0.06} \\
&Fusion Summation & \ranktwo{14.87} & \ranktwo{0.81} & \ranktwo{31.53} & 2.28 & 1.49 & 0.045 \\
&Fusion Attention & 12.36 & 0.7 & 15.45 & 1.22 & \ranktwo{1.266} & 0.003 \\
&Mixture Conventional & 13.03 & 0.73 & 27.18 & 1.84 & 1.484 & 0.049 \\
&Mixture Decoupled & \rankone{18.18} & \rankone{0.93} & \rankone{43.49} & \rankone{3.03} & 1.403 & \rankone{0.065} \\
    \hline
  \end{tabular}
  }
  \label{tab:appendix_em_finetune}
\end{table*}

\begin{figure}[hb]
  \begin{center}
    \includegraphics[width=1.\textwidth]{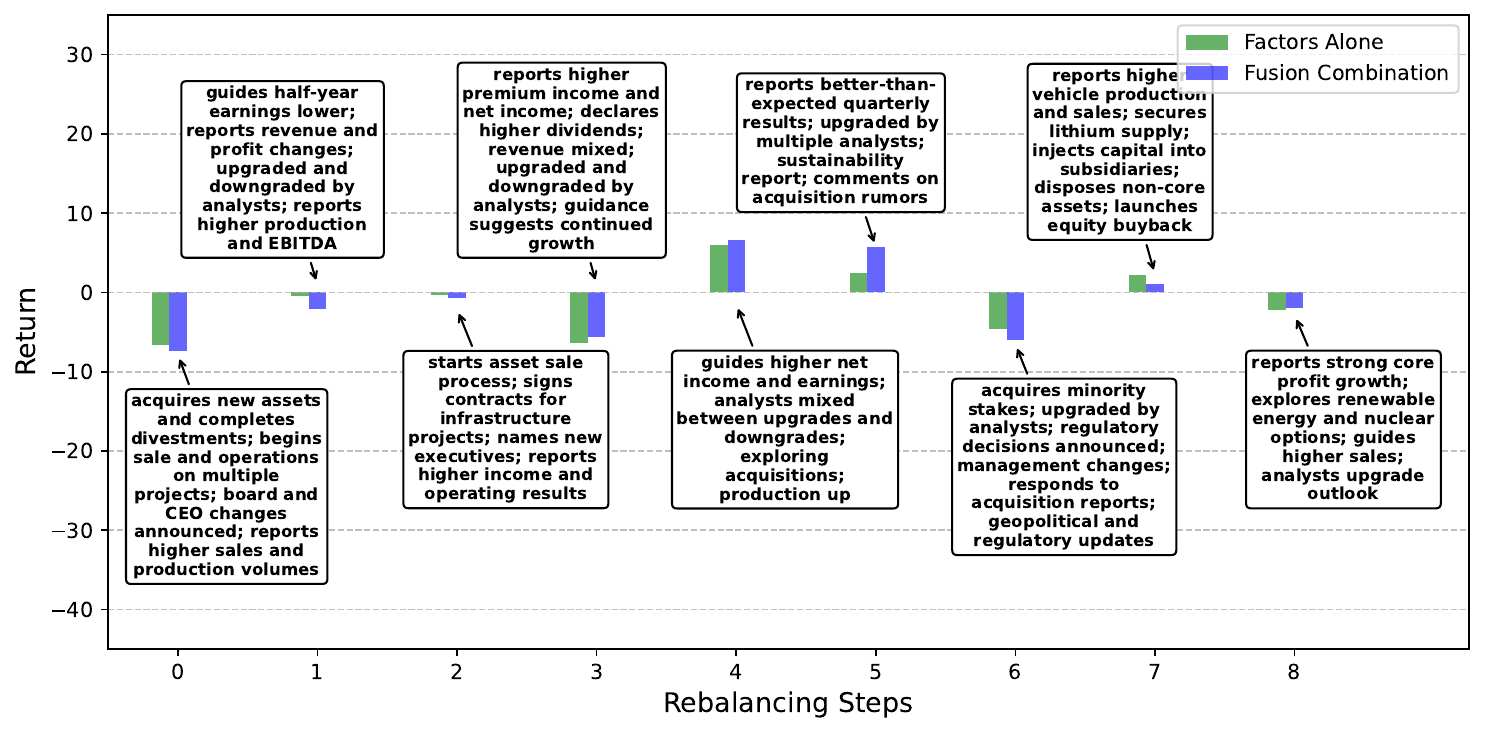}
  \end{center}
  \caption{
  Qualitative Illustration of News Relevance at Rebalancing Steps
  in the Emerging Markets Universe.
  At a given rebalancing step, the newsflow associated with stocks in the long-only portfolio, built based on return predictions, is collected and summarized for visualization in text boxes. 
  The accompanying bar chart presents the forward returns of portfolios constructed using predictions from the \textit{Factors Alone} and \textit{Fusion Combination} methods at each rebalancing step.
  News contributes positively when it provides relevant and complementary information beyond what is captured by quantitative factors. 
  Conversely, it can be detrimental when the information from the news is irrelevant or already reflected in the factors.
  \\
  \textit{Observations:}
  For instance, at steps 0, 1, and 2, news predominantly concerns earnings, sales, ratings, and management changes.
  Such information is likely already priced in factors, causing \textit{Fusion Combination} to struggle to distinguish redundant information from news and lag behind that of \textit{Factors Alone}.
  At steps 3, 4, and 5, in addition to guidance-, income-, and ratings-related news, acquisitions and production updates are also reported, potentially providing complementary information and enhancing the performance of \textit{Fusion Combination}.
  }
  \label{fig:appendix_em_visual_text}
\end{figure}


\clearpage
\subsection{Results of the European Universe}

\begin{table*}[!htbp]
  \centering
  \caption{
  Portfolio and prediction performance of the European Universe.
  The best and second-best results are highlighted with dark gray \rankone{\,\,} and light gray \ranktwo{\,\,} boxes, respectively.
  }
  \resizebox{1.\textwidth}{!}{
  \begin{tabular}{|c|c|c|c|c|c|c|}
    \hline
    \multirow{2}{*}{} & \multicolumn{2}{|c|}{Long-only Portfolios} & \multicolumn{2}{|c|}{Long-short Portfolios} & \multicolumn{2}{|c|}{Prediction Metrics}\\
    \cline{2-7}
    & Ann. Return \% ($\uparrow$) & Sharpe Ratio ($\uparrow$) & Ann. Return \% ($\uparrow$) & Sharpe Ratio ($\uparrow$) & MAPE ($\downarrow$) & IC ($\uparrow$) \\
    \hline
Universe & 6.29 & 0.62 & $-$  & $-$    & $-$ & $-$  \\
Factors Alone & 14.80 & 1.30 & 28.34 & 1.46 & 1.662 & 0.049 \\
News Alone & 9.58 & 0.72 & 1.56 & 0.21 & \rankone{1.111} & 0.001 \\
FININ & 17.01 & 1.32 & 24.54 & 1.41 & 1.316 & 0.051 \\
\hline
Fusion Combination & \rankone{19.6} & \rankone{1.54} & \rankone{32.51} & \rankone{1.70} & 1.302 & \ranktwo{0.052} \\
Fusion Summation & 16.83 & 1.31 & 25.57 & 1.43 & 1.314 & \ranktwo{0.052} \\
Fusion Attention & 14.81 & 1.18 & 20.79 & 1.35 & \ranktwo{1.200} & 0.048 \\
\hline
Mixture Conventional & 16.0 & 1.24 & 25.80 & 1.48 & 1.336 & 0.044 \\
Mixture Decoupled & \ranktwo{18.32} & \ranktwo{1.39} & \ranktwo{30.43} & \rankone{1.70} & 1.318 & \rankone{0.053} \\
  \hline
  \end{tabular}
  }
  \label{tab:appendix_eu_portfolio}
\end{table*}

\begin{figure}[!htbp]
  \begin{center}
    \includegraphics[width=1.0\textwidth]{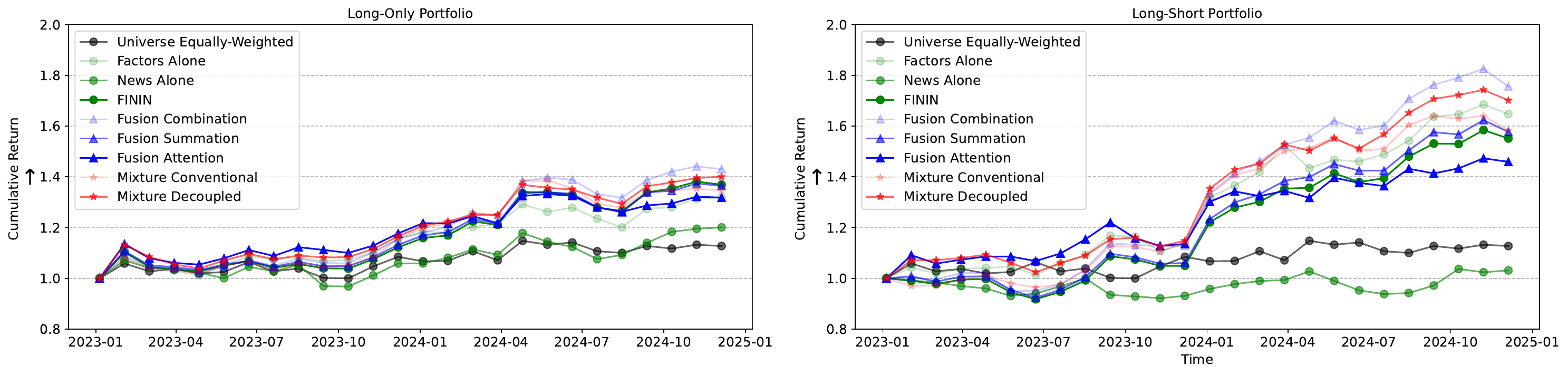}
  \end{center}
  \caption{
  Portfolio Performance Charts of the European Universe.
  }
  \label{fig:eu_charts}
\end{figure}

\begin{figure}[!htbp]
  \begin{center}
    \includegraphics[width=0.7\textwidth]{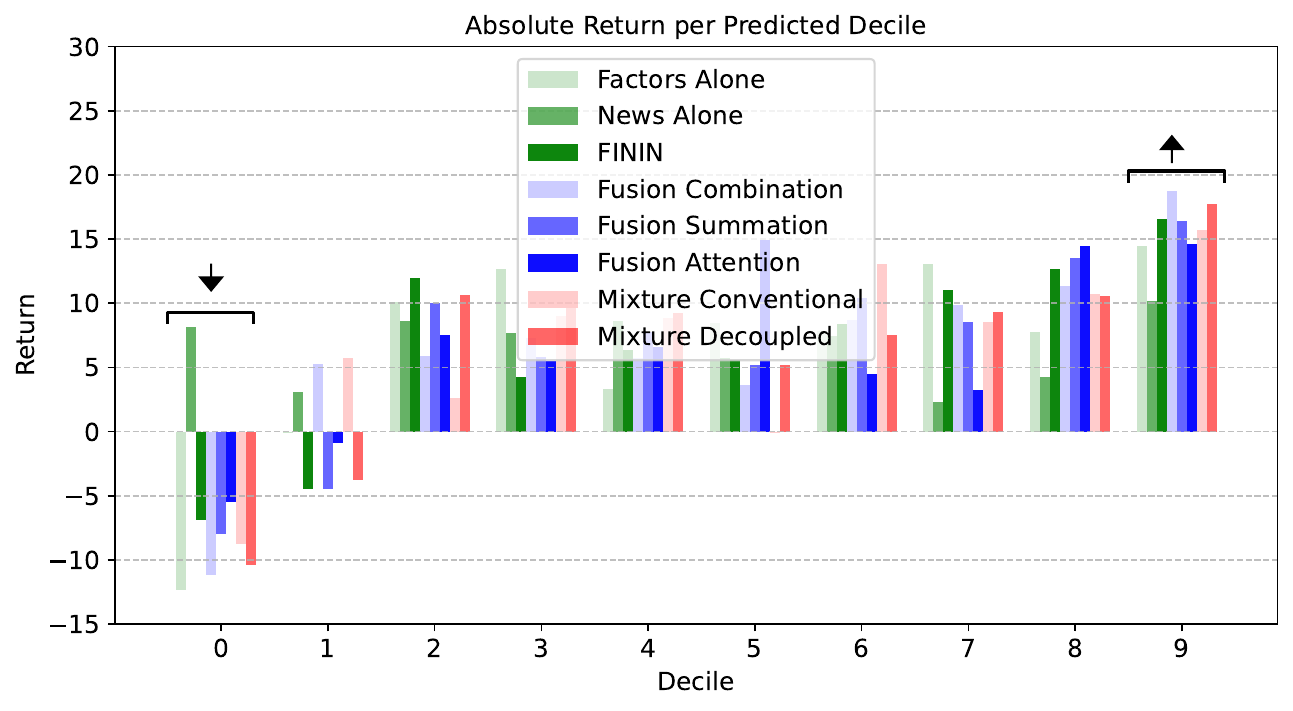}
  \end{center}
  \caption{
  Decile Returns of the European Universe.
  The arrows on the 0th and 9th deciles indicate the desired direction of values.
  A lower return is preferred for the 0th decile, as it represents the short leg of a long-short portfolio.
  }
  \label{fig:appendix_eu_deciles}
\end{figure}

\begin{table*}[!htbp]
  \centering
  \caption{
  Portfolio and prediction performance without and with enabling LLM fine-tuning during training for the European Universe.
  The best and second-best results within each group (without fine-tuning and with fine-tuning) are highlighted with dark gray \rankone{\,\,} and light gray \ranktwo{\,\,} boxes. 
  }
  \resizebox{1.\textwidth}{!}{
  \begin{tabular}{|c|c|c|c|c|c|c|c|}
    \hline
    \multirow{2}{*}{} &  & \multicolumn{2}{|c|}{Long-only Portfolios} & \multicolumn{2}{|c|}{Long-short Portfolios} & \multicolumn{2}{|c|}{Prediction Metrics}\\
    \cline{3-8}
    & & Ann. Return \% ($\uparrow$) & Sharpe Ratio ($\uparrow$) & Ann. Return \% ($\uparrow$) & Sharpe Ratio ($\uparrow$) & MAPE ($\downarrow$) & IC ($\uparrow$) \\
    \hline
    \multirow{7}{*}{\rotatebox{90}{w/o Fine-tuning}}
&News Alone & 9.58 & 0.72 & 1.56 & 0.21 & \rankone{1.111} & 0.001 \\
&FININ & 17.01 & 1.32 & 24.54 & 1.41 & 1.316 & 0.051 \\
&Fusion Combination & \rankone{19.60} & \rankone{1.54} & \rankone{32.51} & \rankone{1.70} & 1.302 & \ranktwo{0.052} \\
&Fusion Summation & 16.83 & 1.31 & 25.57 & 1.43 & 1.314 & \ranktwo{0.052} \\
&Fusion Attention & 14.81 & 1.18 & 20.79 & 1.35 & \ranktwo{1.200} & 0.048 \\
&Mixture Conventional & 16.0 & 1.24 & 25.8 & 1.48 & 1.336 & 0.044 \\
&Mixture Decoupled & \ranktwo{18.32} & \ranktwo{1.39} & \ranktwo{30.43} & \rankone{1.70} & 1.318 & \rankone{0.053} \\
\hline
\multirow{7}{*}{\rotatebox{90}{w/ Fine-tuning}}
&News Alone & 7.54 & 0.74 & -0.03 & 0.03 & \rankone{1.073} & -0.005 \\
&FININ & 18.77 & 1.45 & 29.4 & 1.62 & 1.317 & 0.049 \\
&Fusion Combination & \rankone{20.15} & \ranktwo{1.50} & \rankone{32.27} & \rankone{1.85} & 1.308 & \rankone{0.053} \\
&Fusion Summation & \ranktwo{19.78} & \rankone{1.53} & 30.25 & 1.63 & 1.313 & 0.050 \\
&Fusion Attention & 17.32 & 1.35 & 23.01 & 1.43 & \ranktwo{1.189} & 0.048 \\
&Mixture Conventional & 14.97 & 1.21 & 23.13 & 1.43 & 1.286 & 0.048 \\
&Mixture Decoupled & 18.42 & 1.40 & \ranktwo{31.04} & \ranktwo{1.73} & 1.323 & \rankone{0.053} \\
    \hline
  \end{tabular}
  }
  \label{tab:appendix_eu_finetune}
\end{table*}

\begin{figure}[hb]
  \begin{center}
    \includegraphics[width=1.\textwidth]{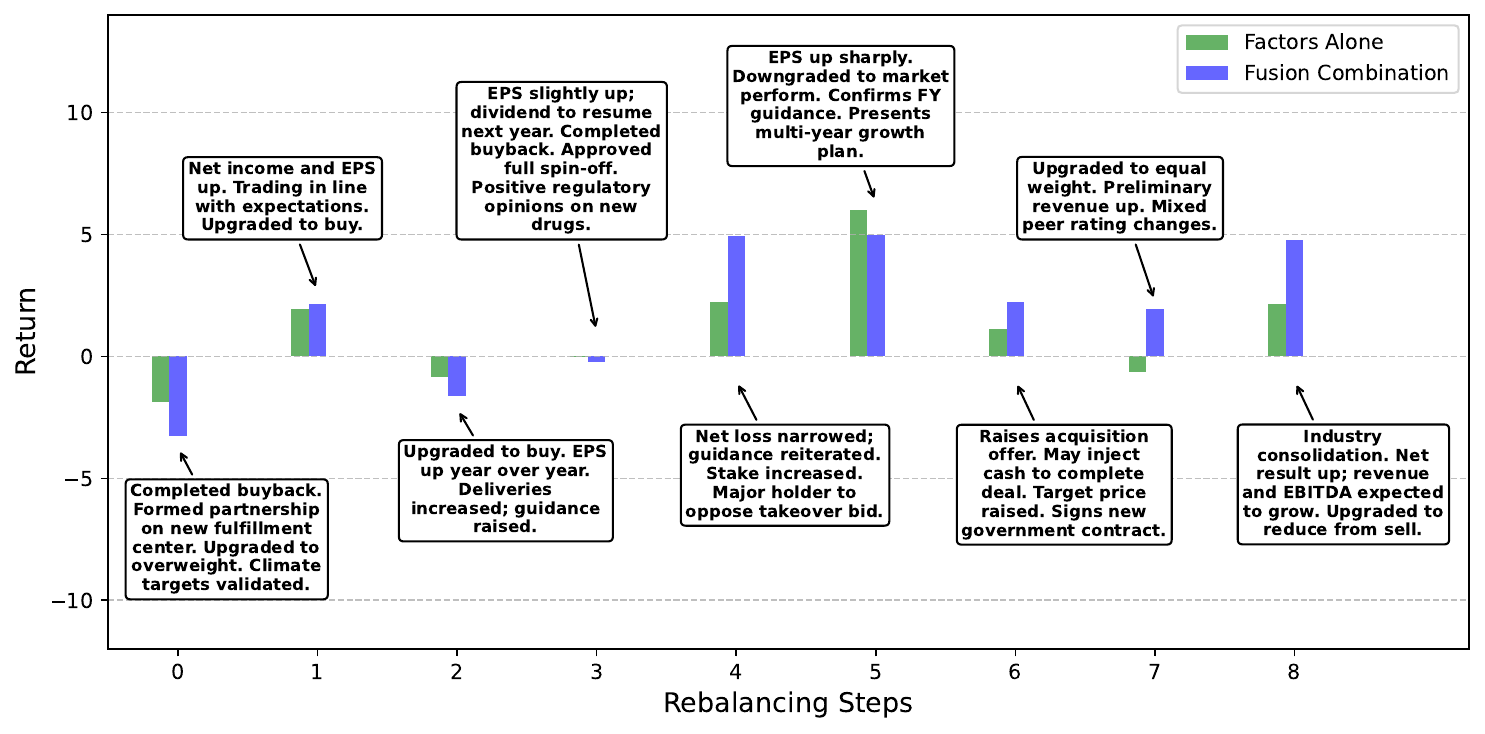}
  \end{center}
  \caption{
  Qualitative Illustration of News Relevance at Rebalancing Steps
  in the European Universe.
  At a given rebalancing step, the newsflow associated with stocks in the long-only portfolio, built based on return predictions, is collected and summarized for visualization in text boxes. 
  The accompanying bar chart presents the forward returns of portfolios constructed using predictions from the \textit{Factors Alone} and \textit{Fusion Combination} methods at each rebalancing step.
  News contributes positively when it provides relevant and complementary information beyond what is captured by quantitative factors. 
  Conversely, it can be detrimental when the information from the news is irrelevant or already reflected in the factors.
  \\
  \textit{Observations:}
  For instance, at steps 0, 2, and 3, news related to earnings, buybacks, and guidance appears to bring little additional information, given the presence of growth and price-based factors in the factors data.
  In these cases, \textit{Fusion Combination} performs worse than \textit{Factors Alone}.
  In contrast, at steps 6, 7, and 8, the news additionally covers new contracts, acquisitions, industry consolidation, and so on, and provides different perspectives that \textit{Fusion Combination} effectively leverages to improve performance.
  }
  \label{fig:appendix_eu_visual_text}
\end{figure}


\end{document}